\documentclass[aps,pra,showpacs,twoside,twocolumn,10pt,superscriptaddress,nofootinbib]{revtex4-1}
\usepackage[colorlinks=true, citecolor=red, urlcolor=blue ]{hyperref}
\usepackage{epsfig,newlfont,amssymb,amsfonts,amsmath,bm,subfigure,palatino,mathtools,amsthm,braket,times,soul,enumitem,color}
\usepackage[normalem]{ulem}
\usepackage{bbold,multirow,soul,graphicx,array}
\usepackage[table,xcdraw]{xcolor}

\newcolumntype{C}[1]{>{\centering\arraybackslash}m{#1}}

\begin{document}
	
	\title{Analytical Percolation Theory for Topological Color Codes under Qubit Loss}
	
	\author{David Amaro}
	\affiliation{Department  of  Physics,  Swansea  University,  Singleton  Park,  Swansea  SA2  8PP,  United  Kingdom.}	
	\author{Jemma Bennett}
	\affiliation{Department  of  Physics,  Swansea  University,  Singleton  Park,  Swansea  SA2  8PP,  United  Kingdom.}	
\affiliation{Department  of  Physics,  Durham  University,  South Road,  Durham DH1 3LE,  United  Kingdom.}	
	\author{Davide Vodola}
		\affiliation{Department  of  Physics,  Swansea  University,  Singleton  Park,  Swansea  SA2  8PP,  United  Kingdom.}
	\author{Markus M{\"u}ller}
		\affiliation{Department  of  Physics,  Swansea  University,  Singleton  Park,  Swansea  SA2  8PP,  United  Kingdom.}

	
	\date{\today}
	
	\begin{abstract}	
		Quantum information theory has shown strong connections with classical statistical physics. For example, quantum error correcting codes like the surface and the color code present a tolerance to qubit loss that is related to the classical percolation threshold of the lattices where the codes are defined. Here we explore such connection to study analytically the tolerance of the color code when the protocol introduced in [Phys. Rev. Lett. \textbf{121}, 060501 (2018)] to correct qubit losses is applied. This protocol is based on the removal  of the lost qubit from the code, a neighboring qubit, and the lattice edges where these two qubits reside. We first obtain analytically the average fraction of edges $ r(p) $ that the protocol erases from the lattice to correct a fraction $ p $ of qubit losses. Then, the threshold $ p_c $ below which the logical information is protected corresponds to the value of $ p $ at which $ r(p) $ equals the bond-percolation threshold of the lattice. Moreover, we prove that the logical information is protected if and only if the set of lost qubits does not include the entire support of any logical operator. The results presented here open a route to an analytical understanding of the effects of qubit losses in topological quantum error codes. 
	\end{abstract}
	
	\maketitle

\section{Introduction}
Quantum information aims to process information by means of quantum systems in order to address problems that are hard to tackle for classical processors. It has shown strong connections with various fields like atomic, molecular and optical (AMO) physics \cite{Ladd2010}, condensed matter \cite{Lewenstein2007,Amico2008}, computer science \cite{Nielsen2000}, and also classical statistical mechanics. The connection between quantum information and classical statistical mechanics has proven to be fruitful in both directions \cite{Cuevas2013,Chubb2019,Zarei2019}. On the one hand a connection between measurement-based quantum computation and classical spin models has been used to show that the partition function of the 2D Ising model can generate the partition functions of all classical spin models \cite{Nest2008-04,Cuevas2009,Xu2011,Cuevas2016}. Furthermore, some quantum algorithms have proven to efficiently approximate the partition function of classical spin models \cite{Nest2007,Cuevas2011,Geraci2008,Lidar1997,Somma2007}. On the other hand, problems in quantum information have found a solution through their connection with solvable classical statistical problems, for instance, to determine which quantum circuits can be efficiently simulated classically \cite{Geraci2010}, or to provide the critical loss threshold of topological quantum error correction (QEC) codes.

To date, topological QEC codes represent one of the most promising routes towards fault-tolerant quantum computation~\cite{Browne2014,Terhal2015}. The logical information is encoded in the joint state of multiple qubits, where information can be protected by applying QEC protocols against noise sources that introduce errors. These QEC protocols consist in the extraction of an error syndrome and the consequent application of a correction. Each QEC code has parameter regimes where errors can or can not be corrected and it was shown that the error threshold that separates those phases is related to the critical point of the order/disorder phase-transition of a statistical physics model~\cite{Jahromi2013,Zarei2018}. For instance, the 2D surface code \cite{Kitaev2003} and the color code \cite{Bombin2006,Bombin2007} under computational (single-qubit bit and phase-flip) errors can be mapped to a 2D random-bond Ising model with two-body \cite{Dennis2002} and three-body interactions \cite{Katzgraber2009}, respectively. Under computational errors and faulty stabilizer measurements the surface code maps to a 3D random-plaquette lattice gauge model \cite{Ohno2004}, while the color code maps to a 3D Ising lattice gauge theory \cite{Andrist2011}. In \cite{Chubb2019} the mapping was was recently extended to account for circuit-level noise in the surface code.


Another particularly damaging noise source is the loss of qubits. A qubit is lost when the information encoded in it can no longer be accessed due to the leakage of the qubit population out of the computational space, or due to the actual loss of particles or photons encoding the qubit. From the theoretical point of view, the loss of information carried by the lost qubits is related to the no-cloning theorem \cite{Bennett1997}, and motivated the proposal of holographic QEC codes \cite{Almheiri2015,Pastawski2015}. Here, the correspondence between the AdS and the CFT spacetimes is identified with the encoding of logical qubits into the multipartite state of the physical qubits. Moreover, in the existing experimental platforms for quantum computation, like trapped ions \cite{Brown2016}, photons \cite{Stefanie2015}, cold atoms \cite{Bloch2008}, or superconducting qubits \cite{Clarke2008}, qubit loss comes in various incarnations like leakage from the computational space or the loss of particles hosting qubits from their traps. A number of protocols to remedy the effect of qubit loss have been proposed and put in practice for trapped ions \cite{Sherman2013}, superconducting qubits \cite{Ghosh2013,Galiautdinov2018,Strikis2019,Rol2019}, photons \cite{Ralph2005,Yang2008}, or quantum dots \cite{Mehl2015,Andrews2018,Chan2019}. 

At the level of QEC codes, there are protocols \cite{Grassl1997,Suchara2015} to correct for the erasure channel, an error model where the position of the lost qubits is known. Some protocols \cite{Delfosse2017a,Delfosse2017b} correct the erasure channel by reinitializing the lost qubits in their computational space and then measuring the stabilizers, producing computational errors at known locations. Another approach consists of removing the lost qubits from the lattice and redefining the code space without the removed qubits. For the surface code, this protocol, which also extends to computational errors, was proposed in \cite{Stace2009,Stace2010}. By mapping the loss events to a percolation problem, it was shown that the surface code presents a tolerance against qubit loss of up to $ 50\% $ in the absence of other sources of error. The correction of qubit losses in the color code has the additional difficulty, compared to the surface code, that the lattice must preserve its trivalence and face-colorability after the code space redefinition. The determination of loss tolerance is of a practical importance for actual and future quantum processors as qubit loss is one of the noise sources of the existing physical platforms.

In \cite{Vodola2018} some of us proposed a protocol to correct qubit losses in the color code that achieved a tolerance of the $ 46(1)\% $ and we showed that, similarly to the surface code, the tolerance of the color code to qubit loss is directly related to a generalized percolation process on the lattice of the color code. More recently, a protocol that consists of mapping the color code to the surface code was proposed in \cite{Aloshious2018}. 


In this work we argue that, given that some logical operators span the three so-called \textit{shrunk lattices}, the critical qubit loss rate $ p_c $ below which the logical information is still protected is directly related to the bond-percolation threshold $ r_c $ of the shrunk lattices of the color code. Here $ p_c $ is the critical value of the qubit loss rate $ p $ at which the average fraction of edges erased $ r(p) $ from a shrunk lattice to correct a fraction $ p $ of lost qubits equals the bond-percolation threshold $ r_c $ of of the corresponding shrunk lattice. Then, by obtaining $ r(p) $ analytically, we are able to obtain $ p_c $ analytically by solving $ r(p_c)=r_c $, as is shown in Fig.~\ref{thresholds}. We apply this prescription to the three regular geometries of the color code and corroborate our results with numerical analysis. We also detail an algebraic technique described in \cite{Vodola2018} and apply it to the three lattices in order to obtain their fundamental qubit loss thresholds $ p_f $. As an additional result, we prove that the logical information is preserved by the loss of qubits if and only if the set of qubits removed from the lattice does not contain the support of any logical operator.

The paper is organized as follows. We start in Sec.~\ref{color_code} by introducing some key concepts about color codes and the notation required for the rest of the paper. Then, in Sec.~\ref{sec_protocol} we review the protocol to correct color codes from qubit losses that was proposed in \cite{Vodola2018}, highlight the connection between the tolerance to qubit loss of the color code with this protocol and the percolation of the color code lattice, and provide detail on the computation of the number of edges erased to correct a qubit loss instance with the protocol. In Sec.~\ref{shrunk_percolation} we analytically derive the relation between the average fraction of edges erased $ r(p) $ and the qubit loss rate $ p $. The Sec.~\ref{sec_results} summarizes the results for the three regular geometries of the color code. In Sec.~\ref{algorithm} we provide an explicit recipe to compute $ r(p) $ up to any order in $ p $. Then, in Sec.~\ref{sec_algebraic} we describe in detail the algebraic technique proposed in \cite{Vodola2018} to obtain the fundamental qubit loss rate $ p_f $, and provide the necessary and sufficient conditions for the existence of the logical information under qubit loss. The values of $ p_c $ and $ p_f $ are summarized in Table \ref{T2}. Finally, we end with the conclusions and outlook in Sec.~\ref{conclusions}.


\section{The color code} \label{color_code}
\begin{figure}[t]
	\centering
	\includegraphics[width=\columnwidth,keepaspectratio]{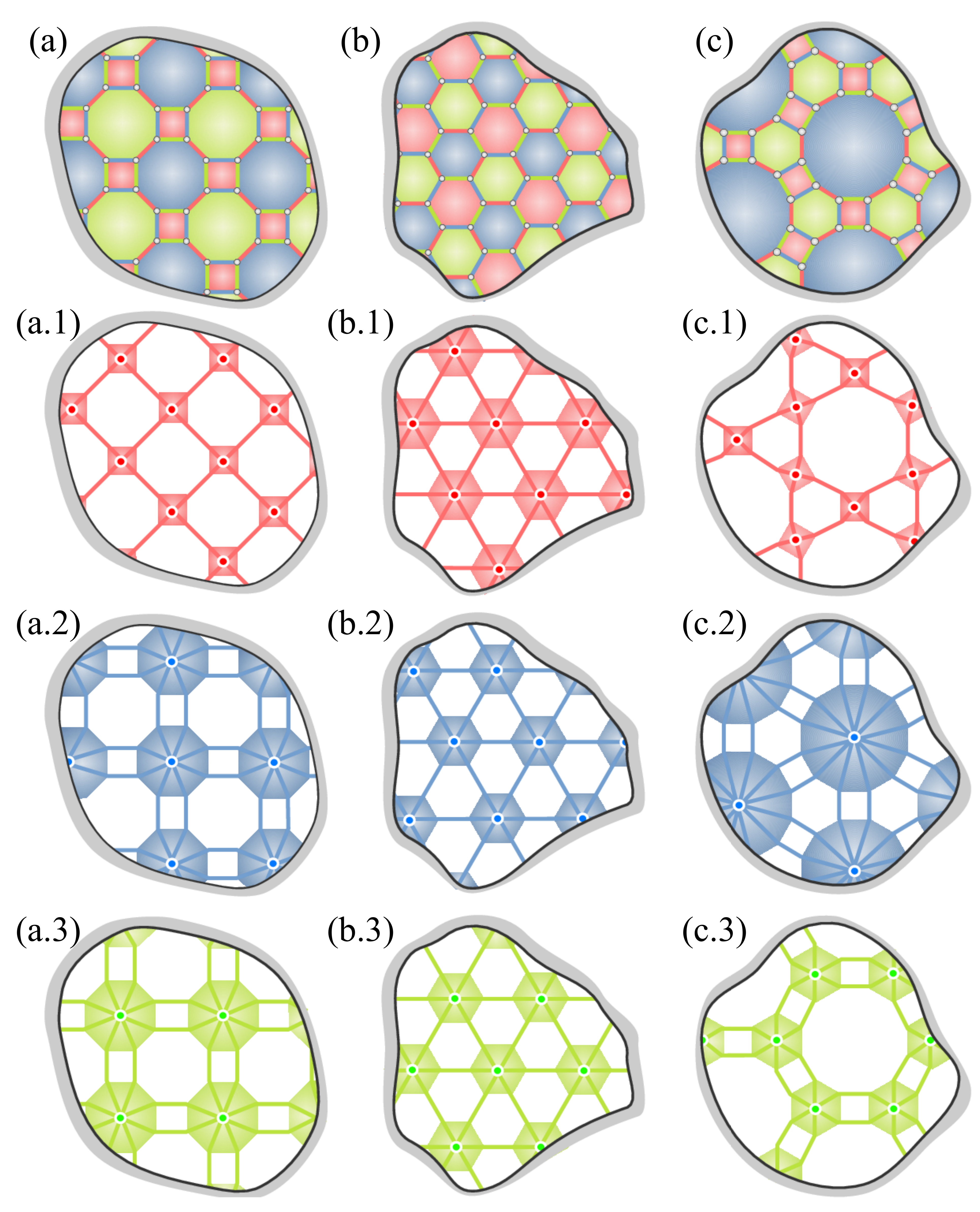}
	\caption{\textbf{Regular geometries of the color code.} Regular trivalent and three-colorable lattices. (a) Lattice 4.8.8 where every node belongs to one square and two octagons. (b) Lattice 6.6.6~(or honeycomb) where every node belongs to three hexagons. (c) Lattice 4.6.12~where every node belongs to one square, one hexagon, and one dodecagon. The red shrunk lattice of the 4.8.8~geometry is (a.1) a square lattice, while the blue (a.2) and green (a.3) shrunk lattices are square lattices with double-bonds. The three shrunk lattices of the 6.6.6~geometry (b.1), (b.2), (b.3) are hexagonal lattices. The red, blue, and green shrunk lattices of the 4.6.12~geometry are (c.1) a kagome lattice, (c.2) a triangular lattice with double-bonds, and (c.3) a hexagonal lattice with double-bonds, respectively.}
	\label{lattices}
\end{figure}

The color code \cite{Bombin2006} is a topological QEC code that protects the logical quantum information by encoding it into a subspace (the code space) of a multi-qubit system. The $ N $ qubits $ i=1,\ldots,N $ sit on the nodes of a trivalent and face-three-colorable lattice. In these lattices, the faces have an even number of nodes, they share two nodes with the adjacent faces, and can be colored with three colors (red, blue, green) such that any two adjacent faces have different color. Similarly, edges can be colored with these three colors such that edges sharing a node have different color, and the color of every edge is different from the color of the faces that it belongs to. The regular lattices that satisfy those properties can be described in vertex notation as a.b.c that indicates that every node in the bulk is shared by three regular polygons with a, b and c vertices. The original and the shrunk lattices of the three regular geometries of the color code, namely the 4.8.8, the 6.6.6, and the 4.6.12~lattices, are depicted in Fig.~\ref{lattices}.

The code space of this stabilizer code~\cite{Gottesman1997} is the common $ +1 $ eigenspace of $ G $ independent and commuting \textit{generators} $ g_{\boldsymbol{f}}^{\sigma} $. A generator is a Pauli operator of type $ \sigma=X,\,Z $ with support on the set of qubits contained by a face of the lattice~$ \boldsymbol{f} $
\begin{equation}
	g_{\boldsymbol{f}}^{\sigma}=\bigotimes_{i\in \boldsymbol{f}}\sigma_i	.
\end{equation}

A code with $ N $ qubits and $ G $ independent generators encodes $ k=N-G $ logical qubits. The $ q $-th logical qubit is defined by two logical generators $ l_{q}^{\sigma} $ for $ \sigma=X,Z $. These operators can be \textit{string operators}, which are defined as
\begin{equation}\label{logical}
l_{q}^{\sigma}=\bigotimes_{i\in \boldsymbol{s}_{q}^{\sigma}}\sigma_i
\end{equation}
on sets of qubits $ \boldsymbol{s}_{q}^{\sigma} $ that take the form of homologically non-trivial strings in the lattice. For example, on the torus, they can be strings wrapping around the ``hole'' and the ``handle''. In a planar code they are strings going from one border to another. 

These strings span the three \textit{shrunk lattices} of the color code. The nodes of the, say, red shrunk lattice are centered on the red plaquettes, and the edges connecting these nodes are the red edges of the color code lattice.

\section{The protocol} \label{sec_protocol}

\begin{figure}[t]
	\centering
	\includegraphics[width=\columnwidth,keepaspectratio]{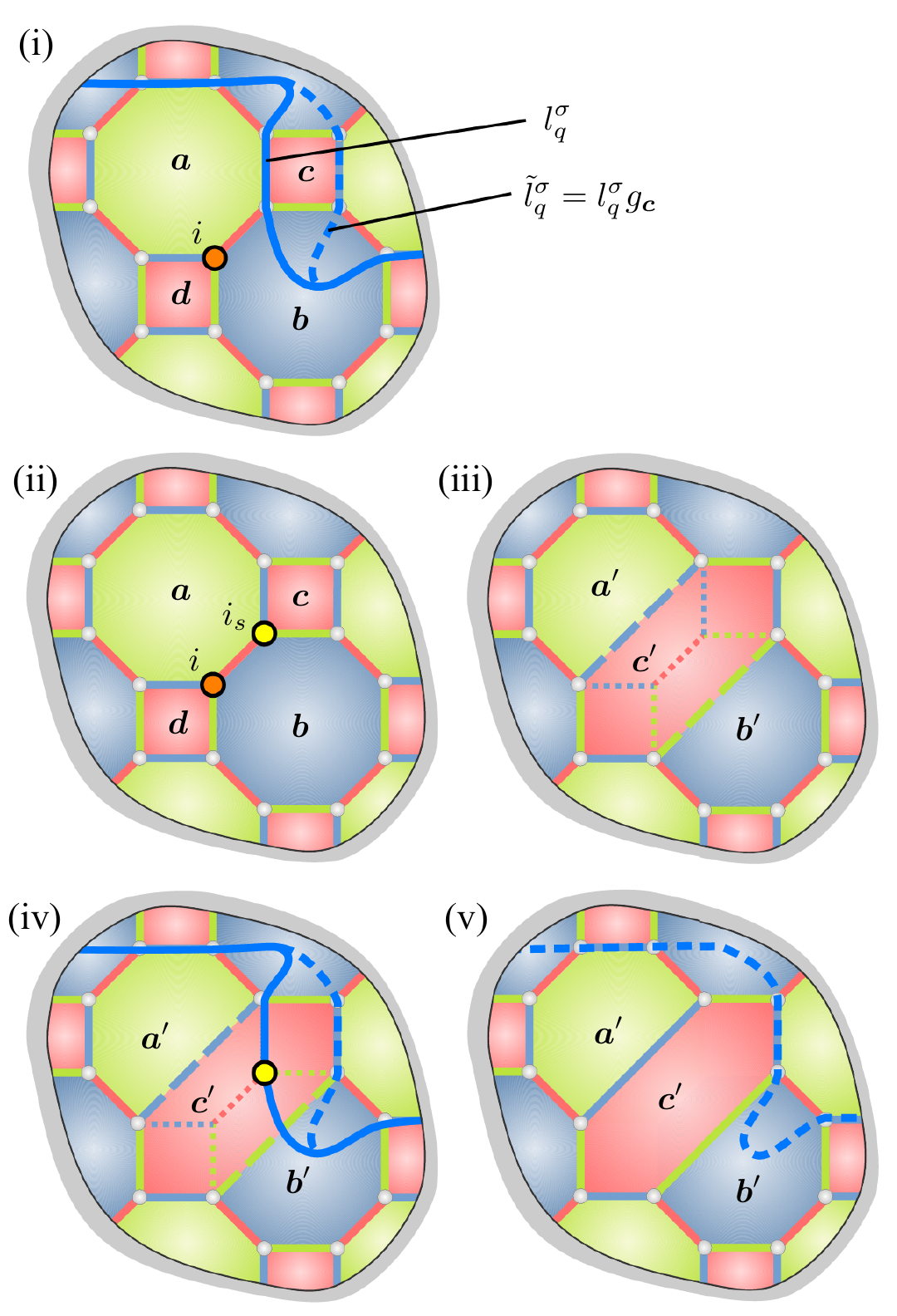}
	\caption{\textbf{Protocol to correct qubit losses on the color code.} (i) Detect the lost qubit $ i $ (orange circle). In this work we assume that the positions of the lost qubits are already known. We also show two string operators $ l_q^\sigma $ (continuous line), and $ \tilde{l}_q^\sigma=l_q^\sigma g_{\boldsymbol{c}}^\sigma $ (dashed line) that differ by multiplication with the generator $ g_{\boldsymbol{c}}^\sigma $ defined on the face $ \boldsymbol{c} $. (ii) Choose a neighboring qubit $ i_s $ as the sacrificed qubit (yellow circle), (iii) remove both $ i $ and $ i_s $ and modify the lattice: the faces $ \boldsymbol{a},\boldsymbol{b} $ that contain both qubits are shrunk into $ \boldsymbol{a}',\boldsymbol{b}' $, and the two faces $ \boldsymbol{c},\boldsymbol{d} $ that contain only one of the removed qubits (lost and sacrificed) are merged into one face $ \boldsymbol{c}' $. This correction erases the five edges adjacent to both qubits (dotted lines) and adds two new edges (dashed lines) such that all remaining qubits have an edge of each color. (iv) Check the existence of the logical information by searching for a well-defined logical operator (like  $ \tilde{l}_q^\sigma $ ) that does not have support on the removed qubits. (v) If the logical information exists, measure the redefined generators $ \boldsymbol{a}',\boldsymbol{b}',\boldsymbol{c}' $. The well defined operators, like $ \tilde{l}_q^\sigma $, remain valid logical operators in the redefined code.}
	\label{fig_protocol}
\end{figure}



The protocol proposed in \cite{Vodola2018} to correct the color code from qubit losses consists in choosing, for every lost qubit,  a neighboring \textit{sacrificed} qubit to be removed together with the loss.  The steps of the protocol are depicted in Fig.~\ref{fig_protocol}. (i) Detect the lost qubits. In this work we will assume that the positions of the lost qubits are known. (ii) Choose the order in which the losses are going to be corrected, and for each loss $ i $, select randomly one of the three neighboring qubits to the loss as the sacrificed qubit $ i_s $. (iii) For each loss, remove the lost qubit and the sacrificed qubit and modify the faces so they do not have support on them: shrink the two faces $ \boldsymbol{a},\,\boldsymbol{b} $ that contain both removed qubits into faces $ \boldsymbol{a}' $ and $ \boldsymbol{b}' $ respectively, and merge the two faces $ \boldsymbol{c},\,\boldsymbol{d} $ that have support on only one of the qubits into a face $ \boldsymbol{c}' $. In this redefinition step the five edges connecting the removed qubits have been erased and two new edges have been added to the lattice. At the same time, a face where two generators are defined is also removed. The new code has two physical qubits and two generators less, so the number of encoded qubits is preserved.

(iv) Check whether the logical information exists or not after the removal of the lost and sacrificed qubits. to this end, a key observation is that logical operators are not uniquely defined. Two logical operators $ l_q^\sigma,\,\tilde{l}_q^\sigma $ belong to the same class $ \{q,\sigma\} $, i.e., they have the same effect on the encoded information, if and only if they differ in a multiplication with a subset $ \mathcal{G} $ of generators
\begin{equation}\label{equivalence}
\tilde{l}_q^\sigma = l_q^\sigma \prod_{g_{\boldsymbol{f}}^{\sigma'}\in \mathcal{G}} g_{\boldsymbol{f}}^{\sigma'}	.
\end{equation}
The logical information still exists in the code if for every class $ \{q,\sigma\} $ there is a well defined logical operator $ \tilde{l}_q^\sigma $, meaning that it does not have support on the removed qubits (lost and sacrificed). For example, in Fig.~\ref{fig_protocol} we show two logical operators that belong to the same class $ \{q,\sigma\} $ because they differ in the multiplication by the generator $ g_{\boldsymbol{c}}^\sigma $: one $ \tilde{l}_q^\sigma $ is well defined, while the other $ l_q^\sigma $ is not. We check the existence of well defined logical operators in two different ways:

(1) Searching in the shrunk lattices for the existence of a \textit{percolating string} without support on the removed qubits. If such strings exists, it corresponds to a logical operator that does not have support on the removed qubits, thus, it is a well defined logical operator. For example, in Fig.~\ref{fig_protocol}(iv) the blue operator $ l_q^\sigma $, which is not well defined, can be deformed into the well defined logical operator $ \tilde{l}_q^\sigma $ by multiplying it with a generator $ g_{\boldsymbol{c}}^\sigma $ of the same type $ \sigma $ but defined on a face of a different color (red face). 
In the same way, finding a percolating string is equivalent to finding a subset of generators $ \mathcal{G} $ such that the logical operator $ \tilde{l}_q^\sigma $ in Eq.~(\ref{equivalence}) does not have support on the removed qubits, with the restriction that these generators have a color different from the color of $ l_q^\sigma $. This method defines the \textit{critical qubit loss rate} $ p_c $ below which the logical information is preserved. The main result of this paper is the analytical computation of $ p_c $ (see Table \ref{T2} for the values obtained), as described in Section \ref{shrunk_percolation}.

\begin{figure}[t]
	\centering
	\includegraphics[width=4.75cm,keepaspectratio]{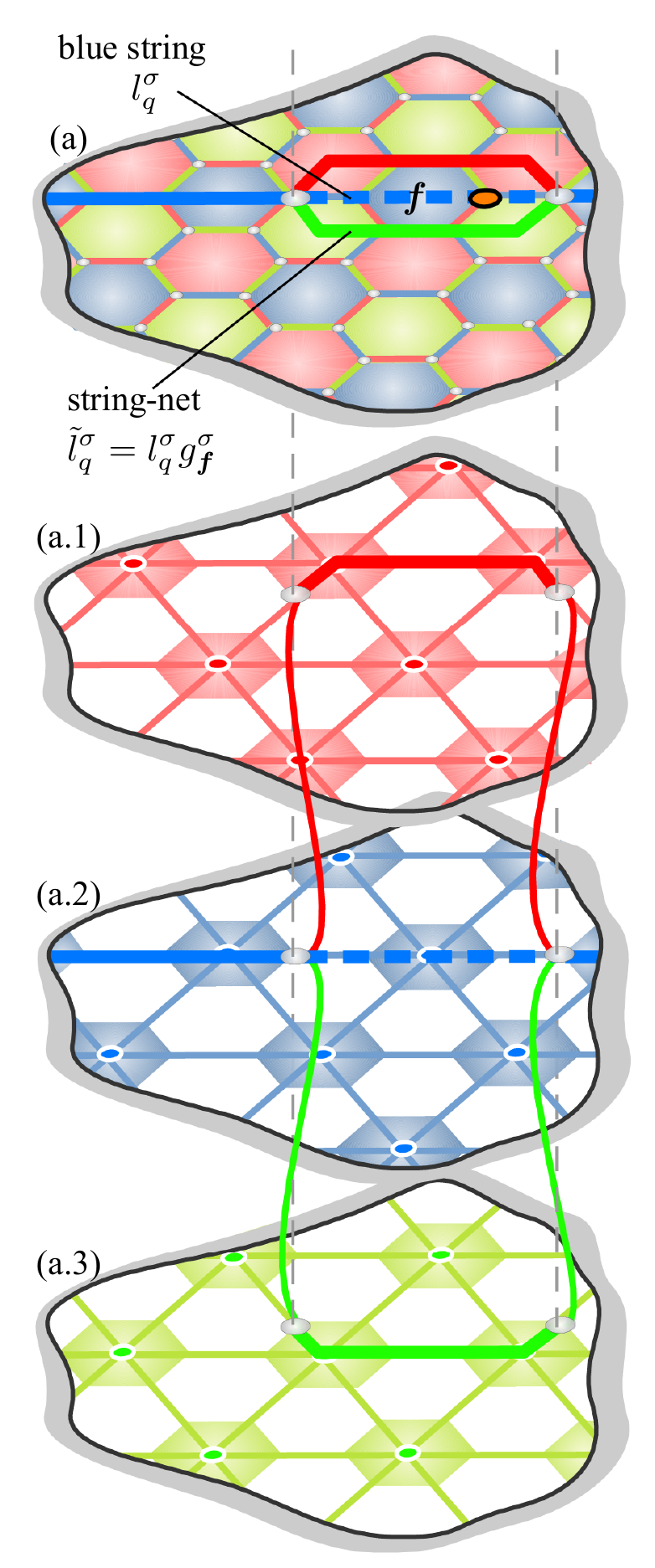}
	\caption{\textbf{Strings and string-nets where logical operators have support.} (a) 6.6.6 color code lattice with a blue string operator $ l_q^\sigma $ on the continuous and discontinuous blue lines, and string-net operator $ \tilde{l}_q^\sigma $. The string-net operator is composed by four paths represented by four continuous lines: (a.1) a red path in the red shrunk lattice, (a.2) two blue paths (the two continuous lines) in the blue shrunk lattice, (a.3) a green path in the green shrunk lattice. Here the blue string operator $ l_q^\sigma $, which is not well defined because it has support on a lost qubit (the orange circle), is multiplied by  the generator $ g_{\boldsymbol{f}}^\sigma $ on the blue face $\boldsymbol{f}$  and transformed into the string-net operator $ \tilde{l}_q^\sigma $ that does not have support on the lost qubit. }	\label{fig_strings}
\end{figure}

(2) The second method consists of directly checking, without any color restriction, the existence of $ \mathcal{G} $ such that $ \tilde{l}_q^\sigma $ in Eq.~(\ref{equivalence}) does not have support on the removed qubits. As this method includes the most general form of a logical operator, it provides the \textit{fundamental threshold} $ p_f $ of the color code affec\-ted by qubit loss (see Table \ref{T2} for the values of $ p_f $ obtained). The solution provided by this method includes in particular the logical operators $ \tilde{l}_q^\sigma $ generated by multiplication with generators of the same color as $ l_q^\sigma $. These logical operators \textit{branch} from one shrunk lattice into the other two, as illustrated in Fig.~\ref{fig_strings}. There a blue string operator, multiplied by a blue generator, branches into the red and the green shrunk lattices and then recombines back to the blue shrunk lattice, taking the form of a \textit{string-net} operator. Therefore, this method is equivalent to a \textit{generalized percolation} problem where the three shrunk lattices are coupled. Despite the exponential number of possible subsets of generators, a solution can be found efficiently, as discussed in Section \ref{sec_algebraic}. Furthermore, in that section we  prove that \textit{given a set of removed qubits $ \boldsymbol{r} $, the logical information is protected if and only if $ \boldsymbol{r} $ does not contain the support of any logical operator.}


(v) If the logical information is preserved, the last step of the protocol consists of projecting the state into the common eigenspace of the redefined generators by generator measurement. As the system is not initially defined in the eigenspace of the redefined generators, excitations may appear when measured, i.e., the system might be projected into the $ -1 $ eigenspace of these generators. These excitations do not need to be removed. Instead, one can define the new code space as determined by the measured eigenvalues of the new generators.


\subsection{Average number of edges erased}\label{corrections}
In order to compute analytically the critical loss rate $ p_c $ at which percolating strings disappear from the shrunk lattices (method (1) of the Sec.~\ref{sec_protocol}), we need to determine the number of edges erased from the original shrunk lattice that we introduce in the following.


\begin{figure}[t]
	\centering
	\includegraphics[width=\columnwidth,keepaspectratio]{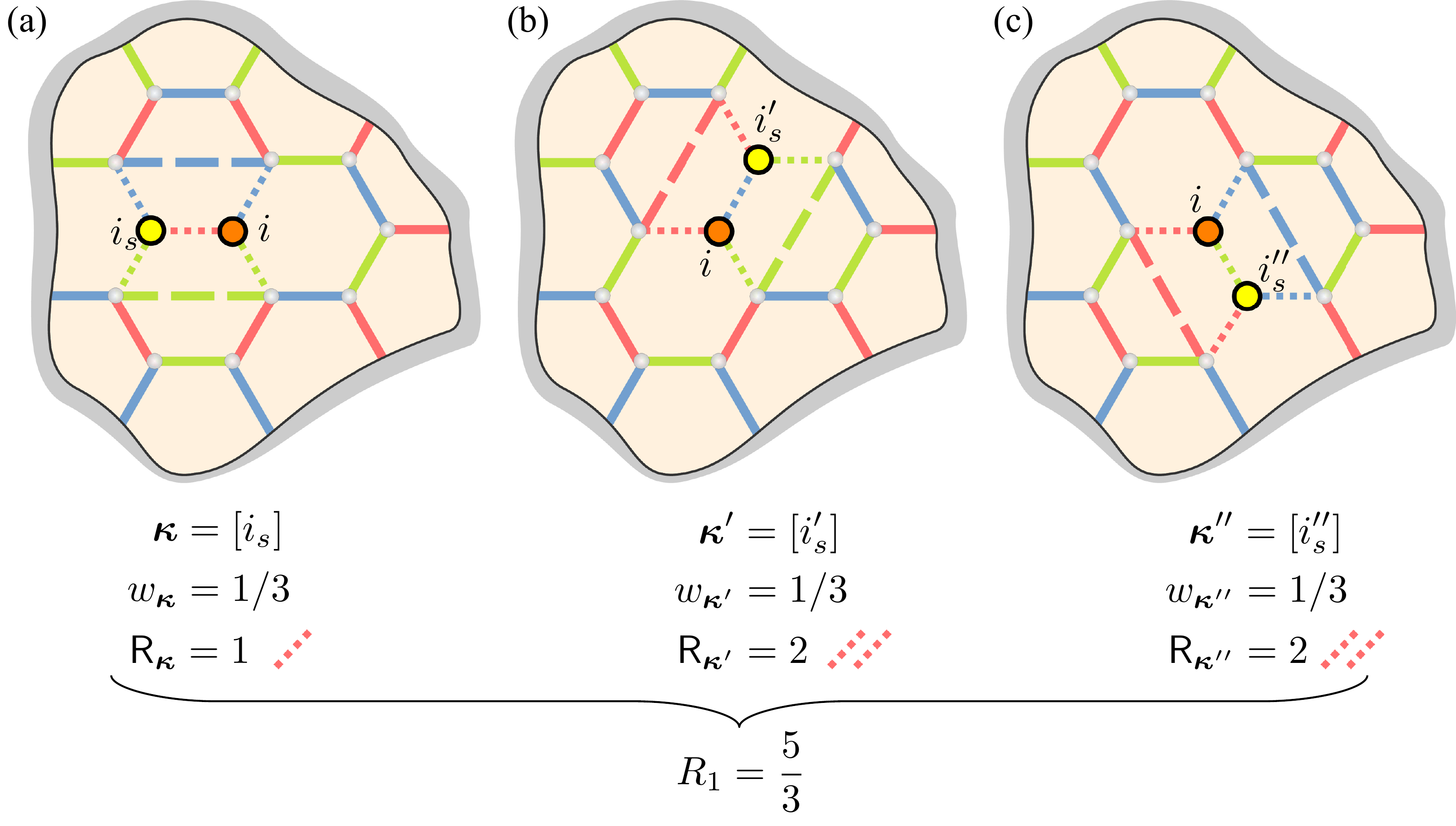}
	\caption{\textbf{One loss corrections.} There are three possible corrections $ \boldsymbol{\kappa} $ for an instance of one qubit loss $ \boldsymbol{i}=\{i\} $ (orange dot) depending on the selection of a neighboring qubit to sacrifice: (a) the qubit $ i_{s} $ on the red edge, (b) the qubit $ i_{s}' $ on the blue edge, (c) the qubit $ i_{s}'' $ on the green edge. We choose each correction with a probability $ w_{\boldsymbol{\kappa}}=1/3 $. From left to right the number of red edges erased (red dotted lines) is $ \mathsf{R}_{\boldsymbol{\kappa}} = 1$, $ \mathsf{R}_{\boldsymbol{\kappa}'}  = 2$, and $ \mathsf{R}_{\boldsymbol{\kappa}''} = 2$. Therefore, the average number of edges erased from the red shrunk lattice by a one-loss event is $ R_1=5/3 $. This value is the same for every loss instance of one qubit loss and for every shrunk lattice.}
	\label{one_loss_correction}
\end{figure} 

Let us define a qubit \textit{loss instance} $ \boldsymbol{i} $ as a set $ \boldsymbol{i}=\{i_1,i_2,\ldots \} $ containing the positions of the $ |\boldsymbol{i}| $ qubits lost. In step (ii) of the protocol, both the order in which qubit losses are corrected, and the sacrificed qubits must be chosen to correct $ \boldsymbol{i} $. In our protocol these selections are made randomly in order to keep the protocol simple and local. Then, every possible \textit{correction} of a loss instance is represented by an ordered list $ \boldsymbol{\kappa}=[i_{s_1},i_{s_2},\ldots] $, where the order corresponds to the order in which the sacrificed qubits $ i_s $ are selected. If we select with equal probability each of the $ |\boldsymbol{i}|! $ orderings and select with equal probability each of the three neighbors of a loss that is corrected, the probability of a correction $ \boldsymbol{\kappa} $ is $ w_{\boldsymbol{\kappa}}=(|\boldsymbol{i}|!)^{-1}3^{-|\boldsymbol{\kappa}|} $, where $ |\boldsymbol{\kappa}| $ is the size of $ \boldsymbol{\kappa} $.

\begin{figure}[t]
	\centering
	\includegraphics[width=\columnwidth,keepaspectratio]{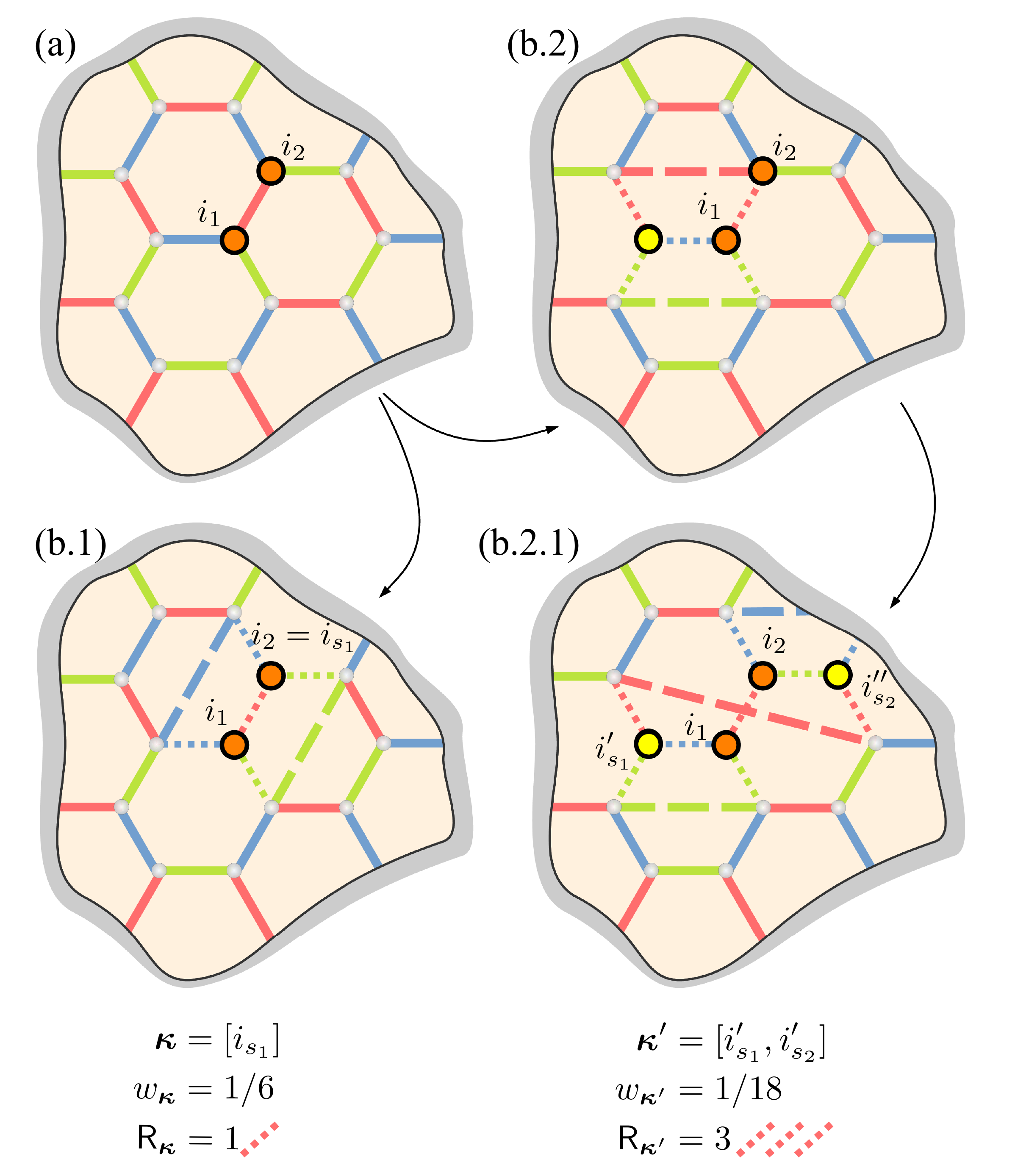}
	\caption{\textbf{Corrections of a loss instance with two qubit losses.} To correct a loss instance $ \boldsymbol{i}=\{i_1,i_2\} $ like the one in (a) composed by two losses indicated with orange dots, the protocol first chooses the order in which the losses are going to be corrected. In this case, the order $ i_1,\,i_2  $ is chosen with a probability of $ 1/2 $. To correct the first loss $ i_1 $, any of the three neighboring qubits can be chosen with a probability $ 1/3 $ as the sacrificed qubit $ i_{s_1} $. In (b.1) the loss $ i_2 $ has been chosen as the sacrificed qubit, so there is no need to correct the loss $ i_2 $. The correction is $ \boldsymbol{\kappa}=[i_{s_1}] $. The probability of this correction is $ w_{\boldsymbol{\kappa}}=(1/2)(1/3)=1/6 $ and $ \mathsf{R}_{\boldsymbol{\kappa}}=1 $ red edges are erased (red dotted lines). In (b.2) a qubit different from the loss $ i_2 $ has been chosen as the sacrificed qubit $ i_{s_1}' $ (yellow dot), and the lattice has been modified accordingly. Then, in (b.2.1) a sacrificed qubit $ i_{s_2}' $ has been chosen to correct the loss $ i_2 $ producing the final erasure of $ \mathsf{R}_{\boldsymbol{\kappa}'}=3 $ red edges with a probability $ w_{\boldsymbol{\kappa}'}=(1/2)(1/3)^2=1/18 $, where the correction is $ \boldsymbol{\kappa}'=[i_{s_1}',i_{s_2}'] $. Note that the new red edge added in (b.2) has not been counted as an erased edge in (b.2.1), because in $ \mathsf{R}_{\boldsymbol{\kappa}} $ we count only those edges erased from the original lattice.}
	\label{two_losses}
\end{figure}

In step (iii) the lattice is modified according to the loss instance $ \boldsymbol{i} $ that occurred and the correction $ \boldsymbol{\kappa} $ selected. In this correction the number of edges erased from the original shrunk lattice is $ \mathsf{R}_{\boldsymbol{\kappa}} $, and the number of edges erased averaged over the set $ \mathcal{K}_{\boldsymbol{i}} $ of all possible corrections of $ \boldsymbol{i} $ is:
\begin{equation}	\label{correction_average}
R_{\boldsymbol{i}} = \sum_{\boldsymbol{\kappa}\in \mathcal{K}_{\boldsymbol{i}}} w_{\boldsymbol{\kappa}} \mathsf{R}_{\boldsymbol{\kappa}}	.
\end{equation}
We notice that, as we are interested in the percolation of the original lattice, in Eq.~\eqref{correction_average} only the links belonging to the original shrunk lattice will be counted. 

As we show in Fig.~\ref{one_loss_correction}, for a loss instance with only one qubit lost $ \boldsymbol{i}=\{i_1\} $, there are three possible corrections $ \boldsymbol{\kappa} $ happening with a probability $ w_{\boldsymbol{\kappa}}=1/3 $, one for every selection of a sacrificed qubit $ i_{s_1} $. The corrections erase $ \mathsf{R}_{\boldsymbol{\kappa}}=1 $, $ \mathsf{R}_{\boldsymbol{\kappa}'}=2 $, and $ \mathsf{R}_{\boldsymbol{\kappa}''}=2 $ red edges, so the average number of edges erased from the original red shrunk lattice to correct $ \{i_1\} $ is:
\begin{equation}
R_1=\frac{5}{3}	.
\end{equation}
$ R_1 $ is the same for every loss instance containing only one loss and it is also the same for every shrunk lattice. Moreover, since every color code is trivalent, $ R_1 $ will be the same for every (also irregular) geometry. 

In Fig.~\ref{two_losses} we show two possible corrections of a two-qubit loss instance $ \boldsymbol{i}=\{i_1,i_2\} $. In the correction depicted in (b.1), the qubit sacrificed $ i_{s_1} $ to correct the loss $ i_1 $ coincides with the second loss $ i_2 $, so no second qubit needs to be sacrificed in order to correct $ i_2 $. The probability of this correction is then $ w_{\boldsymbol{\kappa}}=1/6 $. This correction shows that the set of lost and sacrificed qubits can overlap. In the correction depicted in (b.2.1) two qubits  $ i_{s_1}' $ and $ i_{s_2}' $ have been sacrificed, so the probability is $ w_{\boldsymbol{\kappa}'}=1/18 $. Note that the $ \mathsf{R}_{\boldsymbol{\kappa}'}=3 $ edges erased are counted only from the original shrunk lattice.

\section{Analytical results for percolating strings}\label{shrunk_percolation}
\begin{figure}[t]
	\centering
	\includegraphics[width=\columnwidth,keepaspectratio]{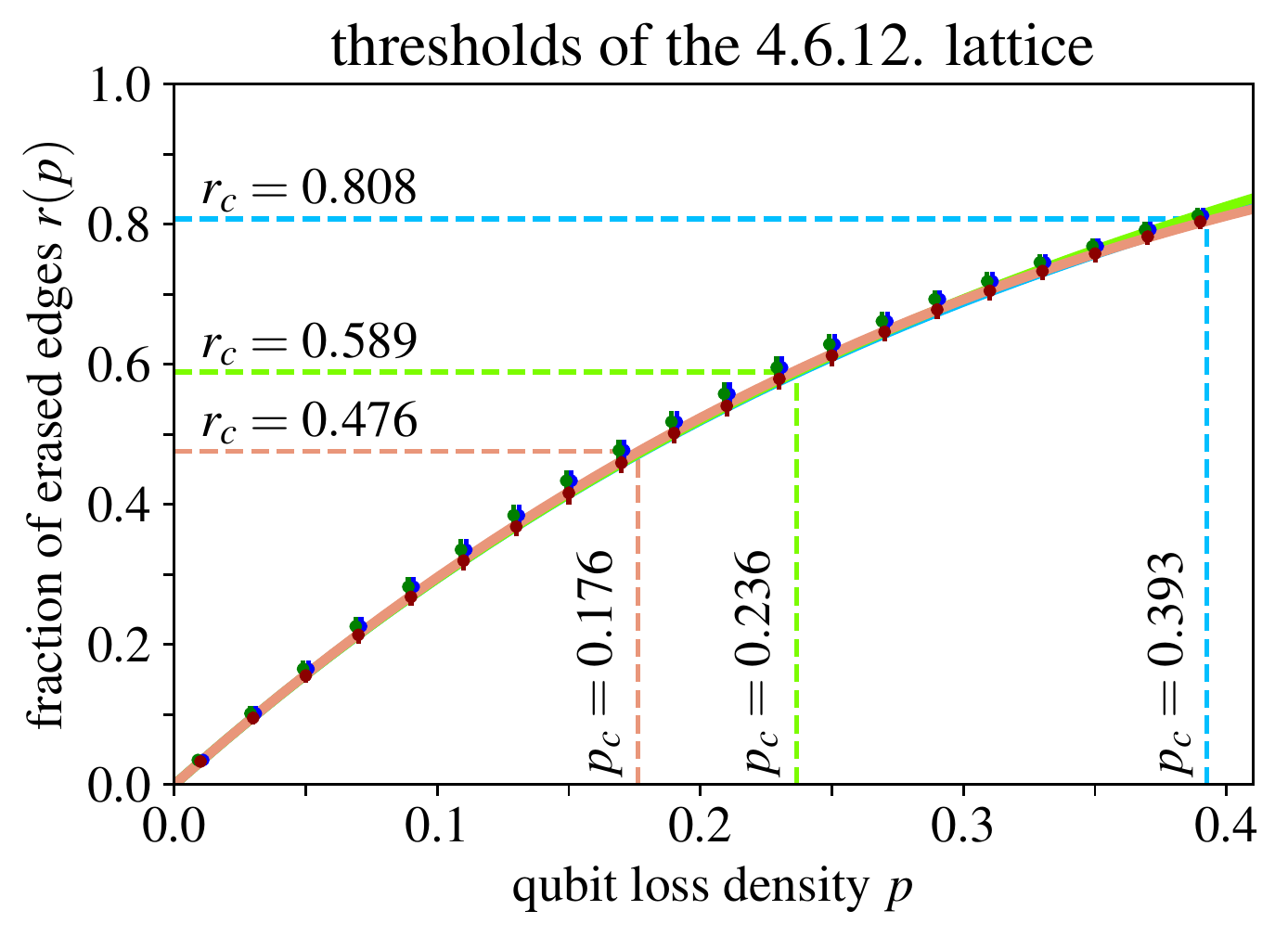}
	\caption{\textbf{Fraction of edges erased} $ r(p) $ as a function of the qubit loss rate $ p $. The points correspond to the numerical estimation, while curves are the analytical estimation. The analytical results with the first three coefficients: $ r(p)\simeq \alpha_1p+\alpha_2p^2+\alpha_3p^3 $ for the red, blue, and green shrunk lattices are represented by the red, blue, and green points and curves respectively. By comparing the analytical curves with the bond-percolation thresholds $ r_c $ taken from Table \ref{T2} we obtain the loss thresholds $ p_c $ of the shrunk lattices. The numerical data is obtained by a Monte-Carlo sampling of losses at various values of the qubit loss rate $ p $.}
	\label{thresholds}
\end{figure}

The main result of this paper is the analytical computation of the critical loss rate $ p_c $ below which there are well defined string operators that percolate through a shrunk lattice. This critical point corresponds to the qubit loss rate $ p $ at which the shrunk lattice does no longer percolate. This happens when the \textit{average fraction of edges erased} $ r(p) $ from the original lattice equals the \textit{bond-percolation threshold} $ r_c $ \cite{Stauffer1985} of the shrunk lattice
\begin{equation} \label{critical}
r(p_c)=r_c	.
\end{equation}
Therefore, $ p_c $ can be obtained analytically from the knowledge of $ r(p) $ and $ r_c $ as shown in Fig.~\ref{thresholds} where we plot the curve $ r(p) $ and the critical loss rates $ p_c $ obtained from the intersection of $ r(p) $ with the values of $ r_c $ for the three shrunk lattices of the 4.6.12 geometry of the color code. In Table \ref{T2} we summarize the values of $ r_c $ and $ p_c $ also for the other geometries.

Note that strings live only on one shrunk lattice, so we can treat the percolation of the three shrunk lattices independently. A value of $ p_c $ is then obtained for each of the three shrunk lattices in each of the three regular geometries of the color code depicted in Fig.~\ref{lattices}.

We study the bond-percolation problem of the shrunk lattice instead of the site-percolation problem because the erased edges of the lattice of the color code coincide with the erased edges of the shrunk lattices, while the removed qubits do not sit on the nodes of the shrunk lattice (recall that the nodes of the shrunk lattices are centered on the plaquettes).

\begin{figure}[t]
	\centering
	\includegraphics[width=\columnwidth,keepaspectratio]{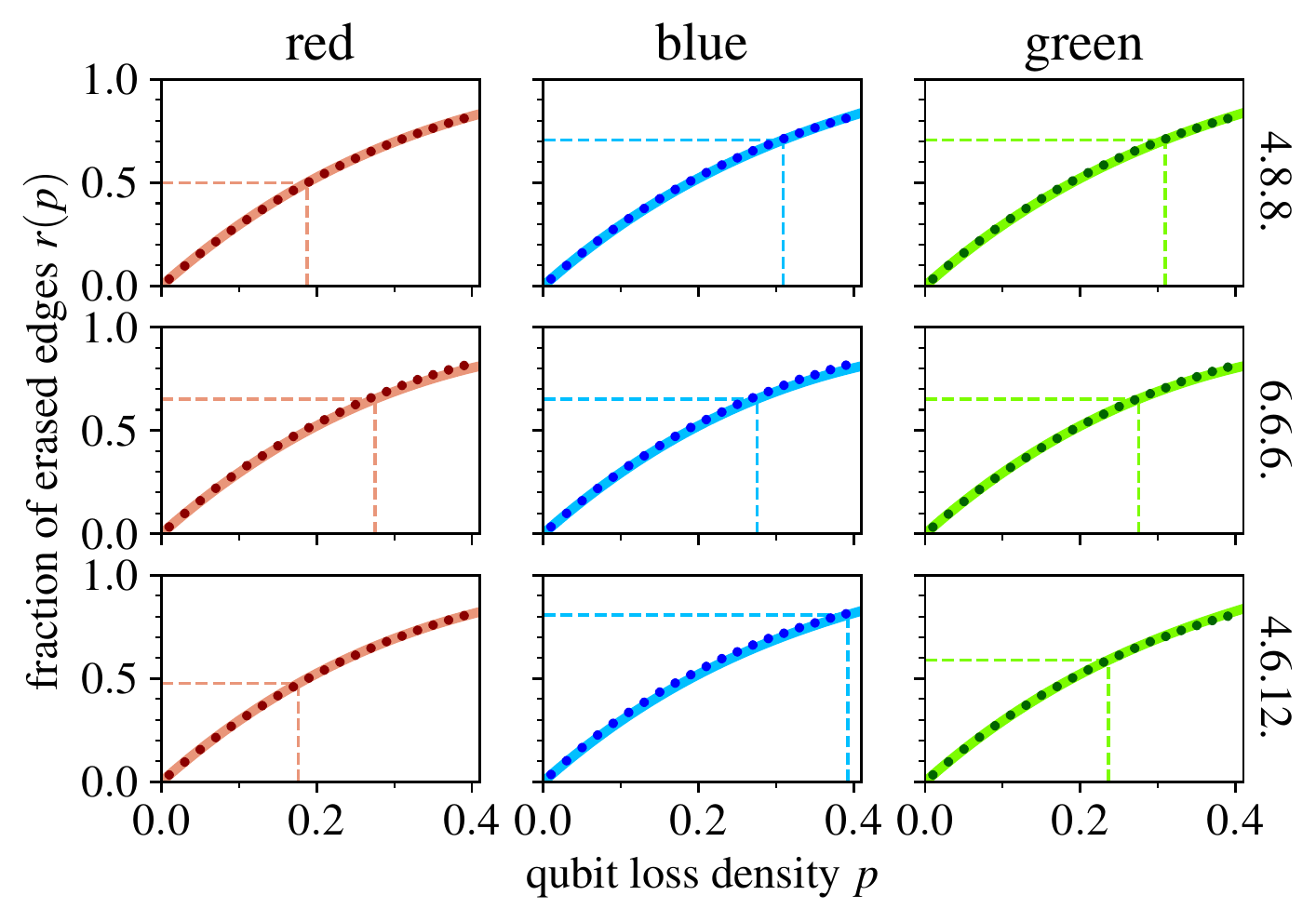}
	\caption{\textbf{Fraction of edges erased} $ r(p) $ as a function of the qubit loss rate $ p $ for every shrunk lattice of the three regular geometries of the color code. The continuous line correspond to the first three orders in the expansion  of $ r(p) $ in powers of $ p $ (Eq.~(\ref{alpha_expansion})). The coefficients of this curve were obtained analytically without performing any approximation. The numerical data (dots) is obtained by a Monte-Carlo sampling of losses at various values of the qubit loss rate $ p $.}
	\label{all}
\end{figure}

We would like to point out that in the bond-percolation problem the edges erased are uniformly distributed in the graph.  However, this is not the case in the color code, given that the edges removed to correct a qubit loss are generally erased in groups, like in Fig.~\ref{one_loss_correction}, where in the last two corrections the two red edges erased are close to each other. However, we assume a uniform distribution of qubit losses without any spatial correlation, so the edges erased will be approximately uniformly distributed, and therefore, we can safely identify $ r_c $ with $ r(p_c) $.

\subsection{Average fraction of edges erased $ r(p) $}\label{power}
The average fraction of edges erased $ r(p) $ is the average number of edges erased divided by the total number of edges $ e=N/2 $ in the shrunk lattice that is being studied, where $ N $ is the total number of qubits. In the following, the error model we consider is the \textit{erasure channel} which assumes local and uncorrelated losses, each of them happening with probability~$ p $. In this noise model $ p $ is also the loss density, so the average number of qubits lost is $ pN $. If the density is low, qubit losses predominantly occur far apart from each other, so they can be treated independently, and therefore, the average number of edges erased by each loss is $ R_1=5/3 $, giving an average fraction of edges erased of $ R_1pN/e=2R_1p $. Then, the average fraction of edges erased grows linearly with $ p $ for low densities:
\begin{equation}	\label{alpha_expansion}
	r(p)=2R_1\,p+\sum_{\ell\geq2}\alpha_\ell\,p^\ell	.
\end{equation}
Our goal is to systematically compute the coefficients $ \alpha_{\ell} $ up to a given desired order $ \ell $. These coefficients are corrections to the linear behavior and they are determined by the \textit{interaction} that takes place between losses that are close to each other. We say that $ \ell $ losses \textit{interact} if the number of edges erased from the original lattice to correct those losses is less than~$ \ell R_1 $, which is the number of edges erased if these losses are far apart from each other. Given that the interaction between losses reduces the number of edges erased, and that the number of interacting instances increases with the density $ p $ of losses, the erasure of edges slows down as $ p $ increases. 
\\

The interaction may come in different fashions as depicted in Fig.~\ref{two_losses}. For example, in the correction (b.1) when the sacrificed qubit coincides with a lost qubit, or in the correction (b.2.1), where one of the edges erased to correct the qubit loss $ i_2 $ is not an edge from the original shrunk lattice but a new edge added from the correction of the first loss $ i_1 $, and therefore, it is not counted in $ r(p) $. If we compute the number of edges erased $ R_{\{i_1,i_2\}} $ for this loss instance as specified by Eq.~(\ref{correction_average}) we will obtain that $ R_{\{i_1,i_2\}} < 2R_1 $.

The interaction between losses can be understood by thinking about the number of edges erased as a sum of \textit{energies}. An instance $ \{i\} $ containing a single loss $ i $ erases an average of $ R_1 $ edges as explained in Fig.~\ref{one_loss_correction}, so let us define $ E_{\{i\}}=R_1 $ as the \textit{internal energy} of every single loss. As mentioned, an instance $ \{i_1,i_2\} $ with two losses erases a number $ R_{\{i_1,i_2\}} $ of edges that might be smaller than $ 2R_1 $, so in this case, there is a non-vanishing interaction energy $ E_{\{i_1,i_2\}} $ that makes $ R_{\{i_1,i_2\}} $ smaller than $ 2R_1 $. We define this two-body interaction energy from the energy sum $ R_{\{i_1,i_2\}}=E_{\{i_1\}}+E_{\{i_2\}}+E_{\{i_1,i_2\}} $. Note that $ E_{\{i_1,i_2\}}=0 $ if the losses do not interact. Analogously, an instance $ \{i_1,i_2,i_3\} $ of three losses erases a number of edges that can be expressed as:
\begin{equation}\label{three_losses_energies}
\begin{split}
	R_{\{i_1,i_2,i_3\}} &= E_{\{i_1\}} +E_{\{i_2\}} +E_{\{i_3\}} + E_{\{i_1,i_2\}} \\
	&+E_{\{i_1,i_3\}}+E_{\{i_2,i_3\}}+E_{\{i_1,i_2,i_3\}}
\end{split}
\end{equation}
where $ \{i_1,i_2\},\,\{i_1,i_3\},\, \{i_2,i_3\} $ are the two-body instances contained in $ \{i_1,i_2,i_3\} $. 

Following this idea, one can write the number of edges erased by any instance as a sum of energies:
\begin{equation}\label{R(E)}
	R_{\boldsymbol{i}} = \sum_{\boldsymbol{j}\subset\boldsymbol{i}} E_{\boldsymbol{j}}
\end{equation}
where the sum is performed over all subsets of the set $ \boldsymbol{i} $. For the empty set $ \emptyset\subset\boldsymbol{i} $ we define the interaction energy $ E_\emptyset=0 $ as zero, while for all the subsets with $ \boldsymbol{j}=\{j\} $ one loss $ j $ the energies are equal: $ E_{\{j\}}=R_1 $. Eq.~\eqref{R(E)} can be represented by a full-rank linear system between $ \{R_{\boldsymbol{i}}\} $ and $ \{E_{\boldsymbol{i}}\} $. By inverting it, we obtain the energies defined by the number of edges erased:
\begin{equation}\label{E(R)}
	E_{\boldsymbol{i}} = (-1)^{|\boldsymbol{i}|}\sum_{\boldsymbol{j}\subset\boldsymbol{i}}(-1)^{|\boldsymbol{j}|}R_{\boldsymbol{j}}
\end{equation}
where $ R_\emptyset=0 $ and $ R_{\boldsymbol{j}}=R_1 $ for all $ \boldsymbol{j} $ with $ |\boldsymbol{j}|=1 $. See Appendix \ref{proof_E(R)} for the proof of this relation. 
\\

Now we can show that the coefficients $ \alpha_{\ell} $ are given by the fully-interacting energies. In our model every loss happens with probability $ p $, so the probability of a loss instance $ \boldsymbol{i} $ is $ p^{|\boldsymbol{i}|}(1-p)^{N-|\boldsymbol{i}|}$. If the average number of edges erased to correct $ \boldsymbol{i} $ is $ R_{\boldsymbol{i}} $, the average fraction of edges erased can be written as:
\begin{equation}\label{fraction_edges_erased}
r(p) = e^{-1}\sum_{\boldsymbol{i}\in\mathcal{I}} p^{|\boldsymbol{i}|}(1-p)^{N-|\boldsymbol{i}|}R_{\boldsymbol{i}}
\end{equation}
where $ \mathcal{I} $ is the set of all possible loss instances. By expanding in powers of $ p $ as done in Appendix \ref{proof_expansion} and using Eq.~(\ref{E(R)}) we can identify the coefficients $ \alpha_{\ell} $ of Eq.~(\ref{alpha_expansion}) with the energies:
\begin{equation}\label{coefficients}
\alpha_\ell = e^{-1}\sum_{\boldsymbol{i}\in\mathcal{I},\;|\boldsymbol{i}|=\ell}E_{\boldsymbol{i}}	.
\end{equation}

However, many energies are zero. For example, as mentioned earlier, the interaction energy of two losses that are far apart from each other vanishes. Analogously, if an instance $ \boldsymbol{i} $ can be split into two disjoint, non-empty subsets $ \boldsymbol{i}^{(A)}\cup \boldsymbol{i}^{(B)}=\boldsymbol{i} $ such that $ R_{\boldsymbol{i}} = R_{\boldsymbol{i}^{(A)}}+R_{\boldsymbol{i}^{(B)}} $ the interaction energy $ E_{\boldsymbol{i}}=0 $ vanishes (proof in Appendix \ref{interacting}), and we call $ \boldsymbol{i} $ a \textit{separable instance}. This happens because the parts $\boldsymbol{i}^{(A)},\boldsymbol{i}^{(B)} $ are too far from each other to interact. On the contrary, the instances that can not be divided in this way are called \textit{fully-interacting} instances, and their energy is non-zero. Therefore the sum over $ \mathcal{I} $ in Eq.~(\ref{coefficients}) can be reduced to the sum over the fully-interacting instances $ \mathcal{I}^{\text{(f-i)}} $.

We also observe that the values of many energies are repeated given that in $ \mathcal{I}^{\text{(f-i)}} $ there are loss instances that are equal up to the symmetries of the lattice of the color code. In the regular geometries of the color code, every node is indistinguishable under the symmetries of the lattice, so we can represent the set of all fully-interacting instances $ \mathcal{I}^{\text{(f-i)}} $ by the set of all fully-interacting instances $ \mathcal{I}^{\text{(f-i)}}_{i_1} $ that have the qubit loss $ i_1 $ in common. Then, every instance $ \boldsymbol{i}\in\mathcal{I}^{\text{(f-i)}}_{i_1} $ is repeated $ N/|\boldsymbol{i}| $ times in $ \mathcal{I}^{\text{(f-i)}} $. Therefore, Eq.~(\ref{coefficients}) can be reduced to:
\begin{equation}\label{coefficients2}
\alpha_\ell = 2\ell^{-1}\sum_{\boldsymbol{i}\in\mathcal{I}^{(\text{f-i})}_{i_1}, \; |\boldsymbol{i}|=\ell}E_{\boldsymbol{i}}  
\end{equation}
where we used that $ e=N/2 $ in the thermodynamic limit. 
      
\begin{figure}[t]
	\centering
	\includegraphics[width=\columnwidth,keepaspectratio]{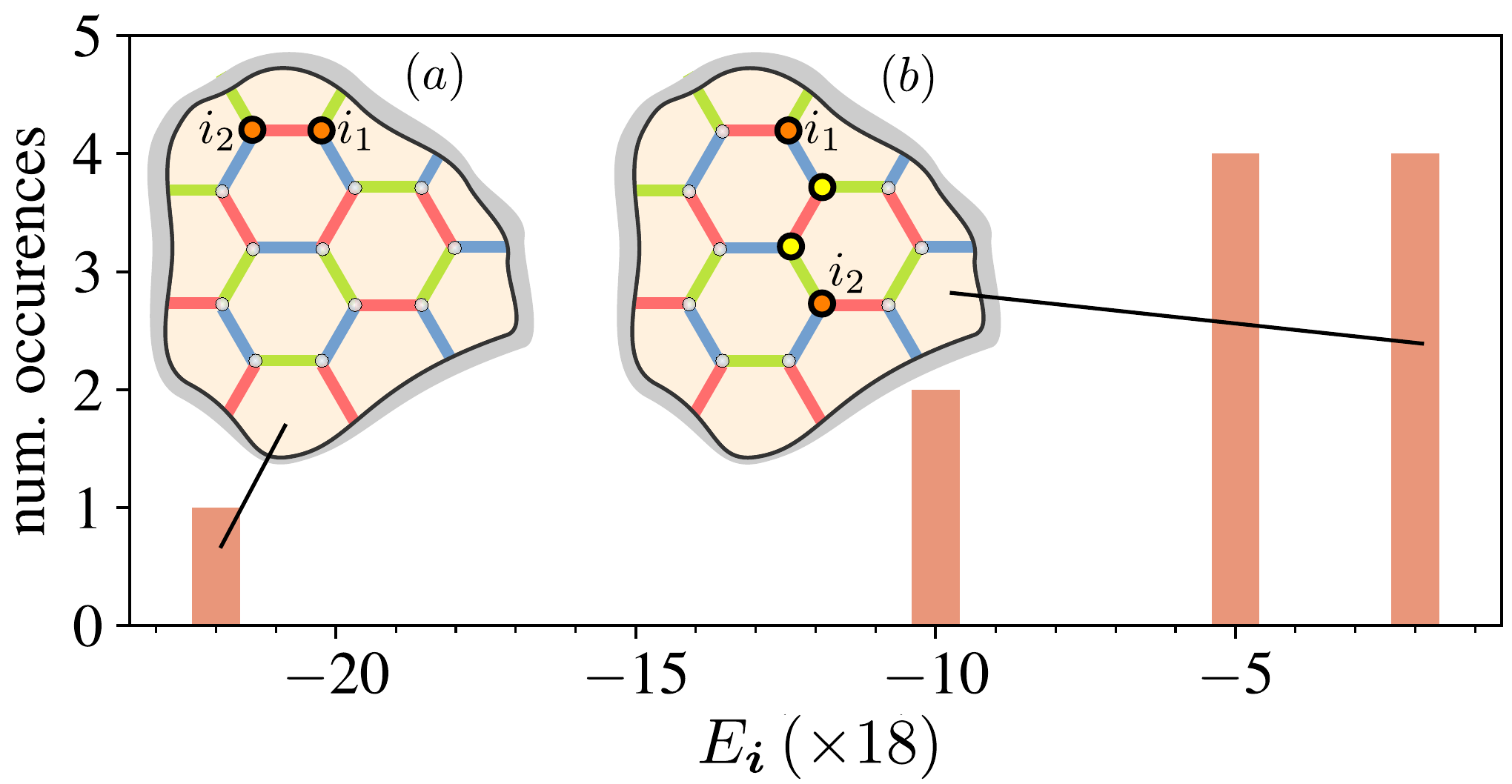}
	\caption{\textbf{Energies $ E_{\boldsymbol{i}} $ of instances} $ \boldsymbol{i}\in\mathcal{I}_{i_1}^{\text{(f-i)}} $ of two losses $ i_1 $, $ i_2 $ for the red shrunk lattice of the 6.6.6~geometry of the color code. In the horizontal axis we indicate the value of the interacting energies computed from the averaged number of edge erased (Eq.~\eqref{E(R)}). These energies are rescaled by a factor of  $ 2! \cdot 3^2=18 $ that represents the number of all possible corrections for each of the two-loss instance. In the vertical axis we indicate the occurrence of each energy, i.e., the number of instances $ \boldsymbol{i}\in\mathcal{I}_{i_1}^{\text{(f-i)}} $ that have the same energy $ E_{\boldsymbol{i}} $. The unique instance that has the biggest energy (in absolute value) is depicted in (a), while one of the four instances with the smallest energy (in absolute value) is depicted in (b). The other three instances with the same energy as (b) can be found by lattice symmetries. The instance in (b) corresponds to an interacting instance since the red link between the two sacrificed qubits (yellow circles) is erased to correct both qubit losses. 
	}
	\label{EvsF}
\end{figure}

For a concrete example, in Fig.~\ref{EvsF}, on the horizontal axis we show the values of the energies $ E_{\boldsymbol{i}} $ of the interacting instances $ \boldsymbol{i} = \{i_1, i_2\} \in\mathcal{I}^{(\text{f-i})}_{i_1} $ and, on the vertical axis, the number of instances that have the same energy. These energies $ E_{\boldsymbol{i}} $ are the ones that appear in Eq.~\eqref{coefficients2}. By recalling that, from Eq.~\eqref{E(R)}, the energy $ E_{\boldsymbol{i}} $ is given by the difference between the number of edges erased by the two-loss instance $\{i_1, i_2\} $ and the number of edges erased separately by each of the single loss $\{i_1\}, \{i_2\} $, it is clear that  the instance that has the biggest energy (in absolute value) corresponds to the couple of qubits residing at the smallest possible distance, as depicted in panel (a). Likewise, the instance that has the smallest energy  (in absolute value) is the one where the qubits have a larger distance that still allows for some corrections to erase a common link (panel (b)). 

Note that to be fully-interacting, all the losses in an instance $ \boldsymbol{i}\in\mathcal{I}^{\text{(f-i)}}_{i_1} $ must be within a finite distance from $ i_1 $. Then, the number of instances in $ \mathcal{I}^{\text{(f-i)}}_{i_1} $ that have up to a certain number of losses $ \ell $ does not depend on the lattice size $ N $. From the number $ I_\ell $ of instances in $ \mathcal{I}^{\text{(f-i)}}_{i_1} $ with $ \ell $ losses we can compute the following averages, that are independent of the system size~$ N $: 
\begin{eqnarray}\label{averages}
\bar{R}_\ell &&= I_\ell^{-1}\sum_{\boldsymbol{i}\in\mathcal{I}^{(\text{f-i})}_{i_1}, \; |\boldsymbol{i}|=\ell} R_{\boldsymbol{i}}\\
\bar{E}_\ell &&= I_\ell^{-1}\sum_{\boldsymbol{i}\in\mathcal{I}^{(\text{f-i})}_{i_1}, \; |\boldsymbol{i}|=\ell} E_{\boldsymbol{i}}
\end{eqnarray}
Note that there is only one instance of one loss, so $ \bar{R}_1=\bar{E}_1=R_1 $. Given that interaction does not increase the number of edges erased, the following hierarchy of inequalities is expected:
\begin{equation}
R_1 \geq \frac{\bar{R}_2}{2} \geq \frac{\bar{R}_3}{3} \geq \cdots \geq \frac{\bar{R}_\ell}{\ell} \geq \cdots \geq \frac{1}{2}	.
\end{equation}

By using these definitions we finally obtain that the coefficients in the power expansion of $ r(p) $ in Eq.~(\ref{alpha_expansion})
\begin{equation}
	\alpha_{\ell}= 2I_\ell\frac{\bar{E}_\ell}{\ell}
\end{equation}
can be seen as the total energy per loss inside the fully-interacting instances. Clearly, given that $ I_\ell $ and $ \bar{E}_\ell $ do not depend on the system size $ N $, the coefficients $ \alpha_{\ell} $ are also independent of the system size. This confirms that the average fraction $ r(p) $ of edges erased from a shrunk lattice depends only on the density of losses $ p $, which is a clear signature of the connection with the percolation theory.

The algorithm that we used to obtain $ I_\ell,\,\bar{R}_\ell,\,\bar{E}_\ell,\,\alpha_\ell $ is described in Section \ref{algorithm}, and the values obtained are summarized in Table \ref{T1}.

\section{Summary of results} \label{sec_results}

We compute the tolerance of the color code under qubit loss in two different ways: (1) searching for percolating strings in the shrunk lattices, and (2) searching for a subset $ \mathcal{G} $ such that the logical operator $ \tilde{l}_q^\sigma $ in Eq.~(\ref{equivalence}) does not have support on the removed qubits. 

Regarding (1) we present the main results of this paper: (1.a) we obtain analytically the average fraction of edges erased $ r(p) $ as a function of the qubit loss rate $ p $, and (1.b) from $ r(p) $ we compute analytically the critical loss rate $ p_c $ below which the logical information is protected. (1.c) We also compare $ r(p) $ with numerical simulations. (1.d) Moreover, $ p_c $ is also computed numerically by an scaling analysis.

\begin{figure}[t]
	\centering
	\includegraphics[width=\columnwidth,keepaspectratio]{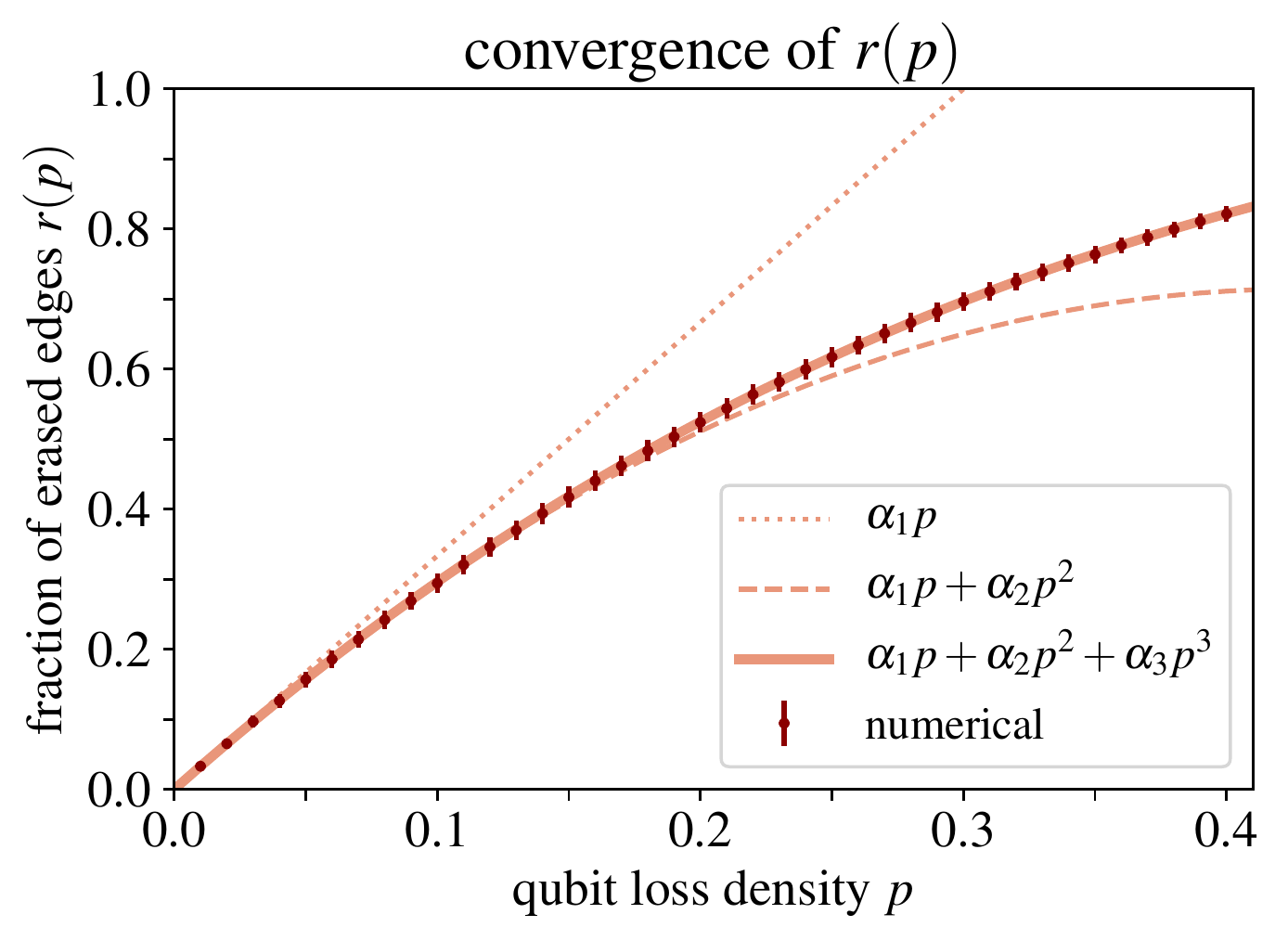}
	\caption{\textbf{Convergence of the first three orders} in the power expansion of the average fraction of edges erased $ r(p) $ for the red shrunk lattice of the 4.8.8~geometry of the color code. We compute analytically the first three coefficients $ \alpha_1,\,\alpha_2,\,\alpha_3 $ in Eq.~(\ref{alpha_expansion}). The dotted line is the first order of the power expansion, the dashed line contains up to the second order, and the continuous line up to the third order. The lines approach the numerical data (red dots) as more orders are added. The numerical data is obtained by a Monte-Carlo sampling of losses at various values of the qubit loss rate $ p $ and a posterior scaling analysis.}
	\label{convergence}
\end{figure} 

\begin{figure}[h]
	\centering
	\includegraphics[width=\columnwidth,keepaspectratio]{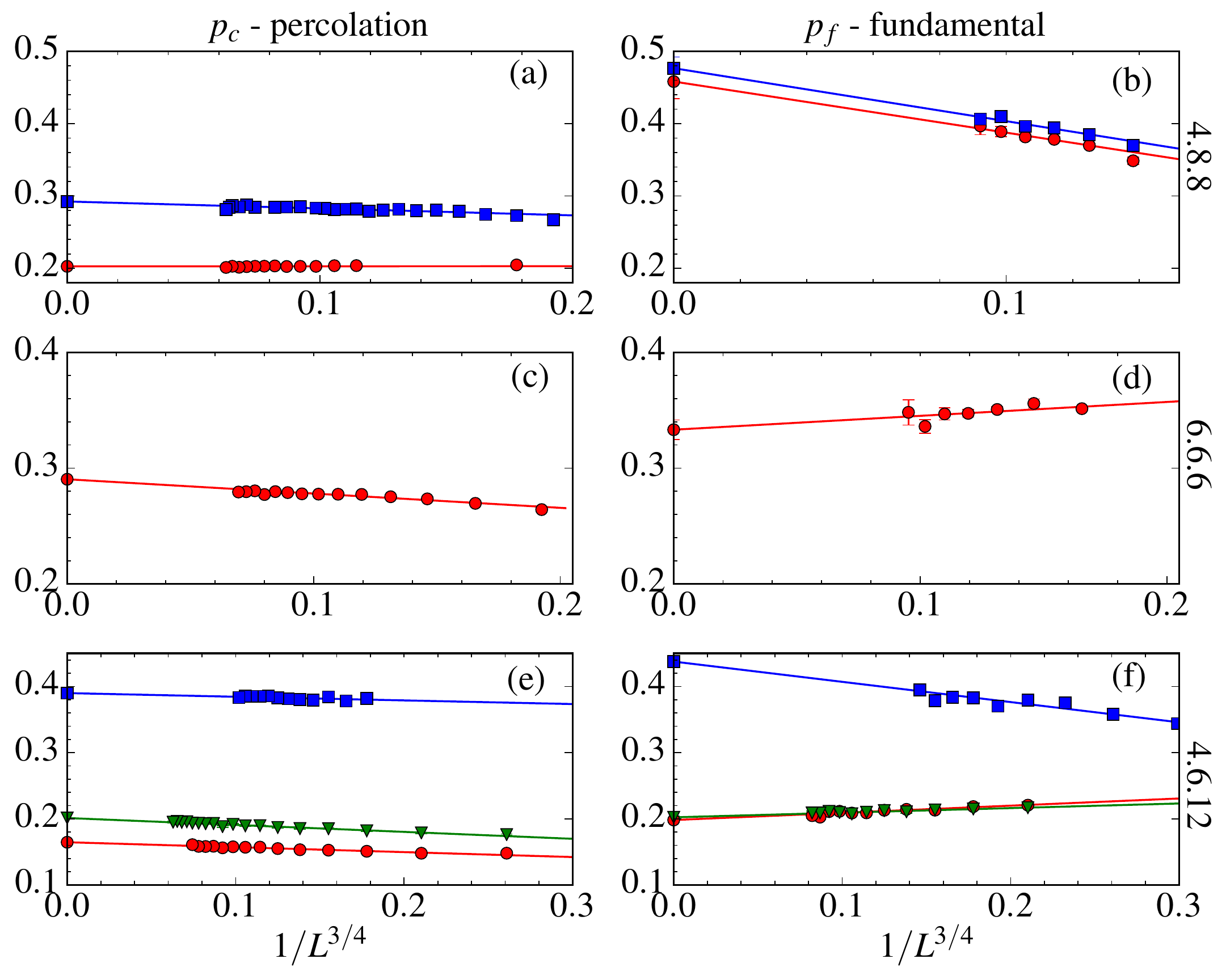}
	\caption{\textbf{Critical qubit loss rate $ p_c $ and fundamental qubit loss rate $ p_f $ obtained numerically.} By sampling loss instances with a Monte Carlo method, we compute the values of $ p_c $ (percolation), and $ p_f $ (fundamental) for various code distances $ L $ of the three regular geometries of the color code. The thresholds are plotted as a function of $1/L^{1/\nu}$ with a critical exponent $\nu = 4/3$ as expected from the percolation theory. Red circles, blue squares and green triangles represent the numerical data for the red, blue and green shrunk lattices, respectively. The continuous lines  fit the points and their intercepts (marked with the same symbols as the data) give the critical threshold in the limit $L\to\infty$. In the graphs (a), (b) for the 4.8.8~lattice, the green shrunk lattice is not represented because it has the same geometry as the blue. In (c), (d) the blue and the green shrunk lattices of the 6.6.6~lattice have the same geometry as the red, so only the red is represented. In (e), (f) for the 4.6.12~lattice, the three shrunk lattices are represented.}
	\label{all_scalings}
\end{figure} 

In relation to (2), we provide in Section \ref{sec_algebraic} an algebraic technique that efficiently finds a solution $ \mathcal{G} $. (2.a) This technique is used in a scaling analysis to obtain numerically the fundamental qubit loss threshold $ p_f $ of the color code. (2.b) Finally we compare the values of $ p_c $ and $ p_f $ obtained.
\\

(1.a) Using the analysis in Section \ref{shrunk_percolation} and the algorithm in Section \ref{algorithm} we compute the first three expansion coefficients $ \alpha_1,\,\alpha_2,\,\alpha_3 $ of $ r(p) $ in Eq.~(\ref{alpha_expansion}) for the three shrunk lattices of the three regular geometries of the color code (values are summarized in Table ~\ref{T1}). Then (1.b), using the bond-percolation thresholds $ r_c $, we obtain $ p_c $ analytically by solving Eq.~(\ref{critical}) up to third order:
\begin{equation}
r_c=\alpha_1p_c+\alpha_2p_c^2+\alpha_3p_c^3	.
\end{equation}
The values of $ r_c $ and $ p_c $ are summarized in Table \ref{T2}. At the critical point $ r(p)\simeq\alpha_1 p+\alpha_2 p^2 +\alpha_3 p^3 $ crosses the value of the bond-percolation threshold $ r_c $ as we show in Figs.~\ref{thresholds} for the 4.6.12~lattice, and in Fig.~\ref{all} for each of the three shrunk lattices of the three regular geometries of the color code. As one can see in Fig.~\ref{thresholds}, the curves $ r(p) $ for the three shrunk lattices of the 4.6.12~color code lattice are almost superposed. Indeed, the curves of all shrunk lattices of all the geometries of the color code depicted in Fig.~\ref{all} are almost superposed (not shown). This indicates that $ r(p) $ does not depend strongly on the geometry of the shrunk lattice. Therefore, the differences between the values of $ p_c $ in the shrunk lattices depend mostly on their bond-percolation threshold $ r_c $. This shows the strong connection between percolation theory and the tolerance of the color code to qubit loss.

\begin{table*}[t]
	\centering
	\renewcommand{\arraystretch}{2}	
	\begin{tabular}{C{5em} | C{4em} C{7em} C{5em} C{6em} || >{\columncolor[HTML]{DCDEFD}}C{6em} >{\columncolor[HTML]{DCDEFD}}C{6em} |>{\columncolor[HTML]{DCDEFD}}C{6em}}
		Geometry& Shrunk	&Geometry			&\multicolumn{2}{c||}{$ r_c  $}					&$ p_c $ an. 	&$ p_c $ num. & $ p_f $	\\
		\hline \hline
		&  Red  	& square 			& $ \frac{1}{2} $	& $ =0.5 $				&$ 0.1877$	& $ 0.2028(7) $ & $ 0.46(1) $			\\
		4.8.8	&  Blue 	& d.b. square 		& $ \sqrt{\frac{1}{2}} $ & $ \simeq0.7071 $		&$ 0.3093$	& $ 0.292(2) $ & $ 0.48(3) $			\\
		&  Green	& d.b. square 		& $ \sqrt{\frac{1}{2}} $ & $ \simeq0.7071 $		&$ 0.3093$	& $ 0.292(2) $ & $ 0.48(3) $			\\
		
		\hline
		&  Red  	& triangular 		& $ 1-2\sin\frac{\pi}{18} $	& $ \simeq 0.6527 $	&$ 0.2752$	& $ 0.290(2) $ & $ 0.33(1) $			\\
		6.6.6	&  Blue 	& triangular 		& $ 1-2\sin\frac{\pi}{18} $	& $ \simeq 0.6527 $	&$ 0.2752$ &	$ 0.290(2) $ & $ 0.33(1) $			\\
		&  Green	& triangular 		& $ 1-2\sin\frac{\pi}{18} $	& $ \simeq 0.6527 $	&$ 0.2752$	& $ 0.290(2) $ & $ 0.33(1) $			\\
		
		\hline
		&  Red  	& kagome 			& \multicolumn{2}{c||}{$ 0.4756 $}				& $ 0.1764$	& $ 0.165(1) $ & $ 0.198(2) $			\\
		4.6.12	&Blue		& d.b. triangular  	& $\sqrt{1-2\sin\frac{\pi}{18}}$ & $ \simeq 0.8079 $ &$ 0.3925$	& $ 0.390(5) $ & $ 0.438(9) $			\\
		&  Green	& d.b. hexagonal 	& $ \sqrt{2\sin\frac{\pi}{18}} $ & $ \simeq 0.5893 $& $ 0.2364$	& $ 0.2012(8) $ & $ 0.202(1) $				
	\end{tabular}
	\caption{\textbf{Tolerance of the color code.} First column: the three regular color code lattices as depicted in Fig.~\ref{lattices}. Second column: their respective shrunk lattices. Third column: geometry of the shrunk lattices (d.b.~stands for double-bonds). Fourth column: analytical and numerical values of the bond-percolation threshold $ r_c $ of the shrunk lattices. In the shrunk lattices with double-bonds, two bonds need to be erased in order to disconnect two nodes. If we call $ r $ the probability of erasure of a bond connecting two nodes, the probability of disconnecting two nodes that are linked via a double bond is $ r^2 $. Therefore, the bond-percolation threshold $ r_c $ (critical probability of disconnecting two nodes) of a shrunk lattice with double-bonds is the square root of the bond-percolation threshold $ \tilde{r}_c $ \cite{Feng2008} of the lattice with simple bonds $ r_c=\sqrt{\tilde{r}_c} $. For shrunk lattices with single bonds $ r_c=\tilde{r}_c $. Fifth column: critical loss threshold $ p_c $ obtained analytically. Sixth column: critical qubit loss rate $ p_c $ obtained by a numerical scaling analysis. Seventh column: fundamental loss threshold $ p_f $ by a numerical scaling analysis. The number between brackets is the error of the last decimal position.}
	\label{T2}
\end{table*}

(1.c) We also estimate $ r(p) $ numerically by performing a Monte Carlo sampling of qubit loss instances for various values of the qubit loss rate $ p $, and estimate the average number of edges erased to correct every instance with a randomly chosen correction. We consider lattices with the three geometries and with a number of qubits close to $ 4000 $. The numerical points obtained are compared with the analytical $ r(p) $  in Figs.~\ref{thresholds} and~\ref{all}. The error bars are comparable with the point size. In the range $ p\in[0.1,0.4] $ that is relevant to obtain $ p_c $ the maximum difference between the analytical (up to third order) and the numerical values of $ r(p) $ is below $ 6\% $. In Fig.~\ref{convergence} we compare the numerical data with the first three orders of $ r(p) $ to show how the curves approximate the numerical data as more expansion terms are added. Limitations of the numerical analysis like the finite-size effects, or the difficulty of sampling instances with a low number of qubits lost are the main sources of discrepancy between the analytical and the numerical analyses.


(1.d) We also obtain $ p_c $ by means of the scaling analysis depicted in the first column of Fig.~\ref{all_scalings} in the following way: In a code of distance $L$, we compute the critical fraction of losses $p_c(L)$ at which, for the first time, a percolating string ceases to exist. It is known that percolation theory predicts \cite{Stauffer1985} the scaling of $ p_c $ as $ L\rightarrow \infty $ to be $ p_c(L)-p_c(\infty)\propto L^{-1/\nu} $, with the scaling exponent $ \nu=4/3 $. This scaling law is followed also by our data. From it, we obtain numerically the value of the critical qubit loss rate $ p_c $ in the thermodynamic limit (when $ L^{-1/\nu}\rightarrow0 $). The values of $ p_c $ obtained numerically by this scaling method are in great accordance with the values obtained by the analytical analysis as can be seen in Table \ref{T2}: the maximum difference is below $ 8\% $.

(2.a) The same scaling analysis is performed in order to obtain the fundamental loss threshold $ p_f $ (second column of Fig.~\ref{all_scalings}). The only difference is that the percolation check is replaced by checking the existence of subset of generators $ \mathcal{G} $ that are a solution of Eq.~(\ref{equivalence}). This subset $ \mathcal{G} $ transforms the original logical operator into a well-defined new logical operator $ \tilde{l}_q^\sigma $ as described in Section \ref{sec_algebraic}. The resulting values of $ p_f $ show the robustness of color codes under qubit loss: for example, the 4.8.8~geometry can tolerate the loss of the $ 46(1)\% $ of the qubits before the first class of logical operators becomes ill defined, which is close to the $ 50 \% $ limit imposed by the non-cloning theorem.

(2.b) The differences between the values of $ p_c $ and $ p_f $, which are easy to visualize in Fig.~\ref{all_scalings}, can be understood by the relation between the two percolation problems that we consider: the percolation of the three decoupled shrunk lattices (provides $ p_c $), and the generalized percolation of the coupled shrunk lattices (provides $ p_f $). Intuitively, $ p_f $ is higher than $ p_c $ because the shrunk lattices with a low bond-percolation threshold $ r_c $ can branch into the other shrunk lattices to increase their tolerance to the erase of edges. For example in the 4.8.8~lattice the red shrunk lattice has a bond-percolation threshold of $ 1/2 $  while the bond-percolation threshold of the blue and the green shrunk lattices is higher: $ 1/\sqrt{2} $. Then, the possibility of branching increases the critical qubit loss rate of the red shrunk lattice of the 4.8.8~geometry from $ p_c\simeq 0.19 $ to the fundamental threshold $ p_f\simeq 0.46 $. On the other hand, given that a shrunk lattice needs the two other lattices to branch, the maximum that $p_f$ can reach is given by the smallest threshold of the other two shrunk lattices. For example, the red shrunk lattice of the 4.6.12~geometry does not improve its tolerance by much (from $ p_c\simeq0.17 $ to $ p_f\simeq0.20 $) by branching into the blue and the green shrunk lattices (despite that the bond-percolation threshold of the blue shrunk lattice is high: $ r_c\simeq0.81 $ and $p_c \simeq 0.39$) because the green shrunk lattice has a low bond-percolation threshold: $ r_c\simeq0.59 $ and $p_c \simeq 0.24$. The relations between $ p_c $ and $ p_f $ for the different shrunk lattices can be easily visualized in Fig.~\ref{all_scalings}.


\begin{table}
	\begin{flushleft}
		\textbf{Input:} Lattice of the color code, a number $ \ell $.\\
		\textbf{Outputs:} Set $ \mathcal{I}^{(\text{f-i})}_{i_1} $ containing all fully-interacting instances $ \boldsymbol{i} $ that have a loss $ i_1 $ in common and contain from $ 2 $ to $ \ell $ losses, the average number of edges erased $ R_{\boldsymbol{i}} $, and the energy $ E_{\boldsymbol{i}} $.
		\begin{enumerate}[label*=\arabic*.]
			\item Place the central loss $ i_1 $ on a qubit of the lattice. Extract the patch $ \boldsymbol{P} $ of qubits at a distance $ 3(\ell-1) $ from $ i_1 $. 
			\item Initialize an empty list $ \mathcal{I}=\{\} $. 
			\item For every instance $ \boldsymbol{i}=\{i,i',\ldots\}\subset\boldsymbol{P} $ containing from $ 2 $ to $ \ell-1 $ losses do:
			\begin{enumerate}[label*=\arabic*.]
				\item Compute $ R_{\boldsymbol{i}} $ with Eq.~(\ref{correction_average}).
				\item Compute $ E_{\boldsymbol{i}} $ with Eq.~(\ref{E(R)}), that requires the value of $ R_{\boldsymbol{i}}$ and the values of $ R_{\boldsymbol{j}} $ with $ j\subset i $ that are stored in $ \mathcal{I} $. Recall that for all instances $ \{i\} $ with only one loss, $ R_{\{i\}}=R_1 $ as explained in Section \ref{corrections}.
				\item Append $ \left[\boldsymbol{i},R_{\boldsymbol{i}},E_{\boldsymbol{i}}\right] $ to $ \mathcal{I}$.
			\end{enumerate}
			\item For every instance $ \boldsymbol{i}=\{i_1,i,i',\ldots\}\subset\boldsymbol{P} $ containing $ \ell $ losses (one of them the central loss $ i_1 $) do:
			\begin{enumerate}[label*=\arabic*.]
				\item Compute $ R_{\boldsymbol{i}} $ with Eq.~(\ref{correction_average}).
				\item Compute $ E_{\boldsymbol{i}} $ with Eq.~(\ref{E(R)}), that requires the value of $ R_{\boldsymbol{i}}$ and the values of $ R_{\boldsymbol{j}} $ with $ j\subset i $ that are stored in $ \mathcal{I} $. Recall that for all instances $ \{i\} $ with only one loss, $ R_{\{i\}}=R_1 $ as explained in Section \ref{corrections}.
				\item Append $ \left[\boldsymbol{i},R_{\boldsymbol{i}},E_{\boldsymbol{i}}\right] $ to $ \mathcal{I}$.
			\end{enumerate}			
			\item Initialize the output list $ \mathcal{I}^{(\text{f-i})}_{i_1}=\{\} $.			
			\item For $ \boldsymbol{i} $ in $ \mathcal{I} $, if $ E_{\boldsymbol{i}}\neq0 $ and $ i_1\in \boldsymbol{i} $, append $ \left[\boldsymbol{i},R_{\boldsymbol{i}},E_{\boldsymbol{i}}\right] $ to $ \mathcal{I}^{(\text{f-i})}_{i_1} $.
			\item Return $ \mathcal{I}^{(\text{f-i})}_{i_1} $.
		\end{enumerate}
	\end{flushleft}
	\caption{Pseudo-code summarizing the main steps to generate all fully-interacting instances $ \boldsymbol{i}\in\mathcal{I}^{(\text{f-i})}_{i_1} $ that contain $ \ell $ losses or less, the average number of edges erased $ R_{\boldsymbol{i}} $, and their energy $ E_{\boldsymbol{i}} $. The coefficients $ \alpha_{\ell} $ in Eq.~(\ref{alpha_expansion}) can be computed from these values with Eq.~(\ref{coefficients2}).}	\label{iterative}
\end{table}

\begin{table*}[t]
	\centering
	\renewcommand{\arraystretch}{2}	
	\begin{tabular}{C{4em} C{3em}|C{2em}C{2em}C{2em}>{\columncolor[HTML]{DCDEFD}}C{2em}|C{2em}C{2em}C{2em}>{\columncolor[HTML]{DCDEFD}}C{2em}|C{2em}C{3em}C{2em}>{\columncolor[HTML]{DCDEFD}}C{2em}}
		Geometry& Shrunk	&$I_1$	&$R_1$	&$E_1$	&$ \alpha_1 $	&	$ I_2 $& $ \bar{R}_2 $		& $ \bar{E}_2 $		&$\alpha_2$		&$ I_3 $& $ \bar{R}_3 $		 & $ \bar{E}_3 $		 & $ \alpha_3 $\\
		\hline \hline
		&  Red  &$ 1 $	&$ \frac{5}{3} $	&$ \frac{5}{3} $	&$ \frac{10}{3} $	& $ 11 $&$ \frac{295}{99} $ &$ -\frac{35}{99} $ &$-\frac{35}{9}$&$ 72 $ &$ \frac{3995}{972}$ &$\frac{35}{972}$ & $ \frac{140}{81} $\\
		4.8.8	&  Blue &$ 1 $	&$ \frac{5}{3} $	&$ \frac{5}{3} $	&$ \frac{10}{3} $	& $ 9 $ &$ \frac{233}{81} $ &$ -\frac{37}{81} $ &$-\frac{37}{9}$&$ 102 $&$\frac{5749}{1377}$ &$\frac{95}{2754}$ & $ \frac{190}{81} $\\
		&  Green&$ 1 $	&$ \frac{5}{3} $	&$ \frac{5}{3} $	&$ \frac{10}{3} $	& $ 9 $	&$ \frac{233}{81} $ &$ -\frac{37}{81} $ &$-\frac{37}{9}$&$ 102 $&$\frac{5749}{1377}$ &$\frac{95}{2754}$ & $ \frac{190}{81} $\\
		
		\hline
		&  Red  &$ 1 $	&$ \frac{5}{3} $	&$ \frac{5}{3} $	&$ \frac{10}{3} $	& $ 11 $&$ \frac{295}{99} $ &$ -\frac{35}{99} $ &$-\frac{35}{9}$& $122 $&$\frac{14161}{3294}$&$\frac{29}{1647}$	& $ \frac{116}{81} $\\
		6.6.6	&  Blue &$ 1 $	&$ \frac{5}{3} $	&$ \frac{5}{3} $	&$ \frac{10}{3} $	& $ 11 $&$ \frac{295}{99} $ &$ -\frac{35}{99} $ &$-\frac{35}{9}$& $122 $&$\frac{14161}{3294}$&$\frac{29}{1647}$	& $ \frac{116}{81} $\\
		&  Green&$ 1 $	&$ \frac{5}{3} $	&$ \frac{5}{3} $	&$ \frac{10}{3} $	& $ 11 $&$ \frac{295}{99} $ &$ -\frac{35}{99} $ &$-\frac{35}{9}$& $122 $&$\frac{14161}{3294}$&$\frac{29}{1647}$	& $ \frac{116}{81} $\\
		
		\hline
		&  Red  &$ 1 $	&$ \frac{5}{3} $	&$ \frac{5}{3} $	&$ \frac{10}{3} $	& $ 11 $&$ \frac{295}{99} $ &$ -\frac{35}{99} $ &$-\frac{35}{9}$&$ 64 $&$ \frac{7057}{1728} $	&$ \frac{1}{27} $ 	 &$ \frac{128}{81} $ \\
		4.6.12	&  Blue &$ 1 $	&$ \frac{5}{3} $	&$ \frac{5}{3} $	&$ \frac{10}{3} $	& $ 9 $ &$ \frac{233}{81} $ &$ -\frac{37}{81} $&$-\frac{37}{9}$ &$ 91 $& $ \frac{10214}{2457} $&$ \frac{89}{2457} $  &$\frac{178}{81}$\\
		&  Green&$ 1 $	&$ \frac{5}{3} $	&$ \frac{5}{3} $	&$ \frac{10}{3} $	& $ 9 $ &$ \frac{233}{81} $ &$ -\frac{37}{81} $ &$-\frac{37}{9}$&$ 102 $ &$ \frac{5749}{1377} $ &$ \frac{95}{2754} $&$\frac{190}{81}$	
	\end{tabular}
	\caption{\textbf{Results for the analytical expansion of $r(p)$}. Representative factors for $ \ell=1,2,3 $ losses for the three shrunk lattices in three regular geometries of the color code. The number of fully-interacting instances is $ I_\ell $, the average number of edges erased by them is $ \bar{R}_\ell $, and the average energy associated is $ \bar{E}_\ell $. The coefficients of the power expansion in Eq.~(\ref{alpha_expansion}) are $ \alpha_\ell $. All these quantities have been obtained analytically without perfoming any approximation.}
	\label{T1}
\end{table*}

\section{Computation of the coefficients $ \alpha_\ell $}	\label{algorithm}
In this Section we provide an algorithm to compute the expansion coefficients $ \alpha_{\ell} $ of $ r(p) $ in Eq.~(\ref{alpha_expansion}). The computation of the first $ \ell $ coefficients as in Eq.~(\ref{coefficients2}) requires the energies $ E_{\boldsymbol{i}} $ of all the fully-interacting loss instances $ \boldsymbol{i}\in\mathcal{I}^{(\text{f-i})}_{i_1} $ that have the loss $ i_1 $ in common and that contain from $ 2 $ to $ \ell $ losses. We explain the algorithm for the case of $ \ell=3 $ losses, and provide the pseudo-code in Table \ref{iterative} for any $ \ell $. The steps of the algorithm are the following:


$ 1. $ Place the central loss $ i_1 $  on a qubit in the lattice and extract a set of qubits  $ \boldsymbol{P} $ (we call it a patch) at a finite distance from $ i_1 $. By the \textit{distance} between two nodes we mean the number of edges in the shortest path that connects these nodes. In order to consider all fully-interacting instances in $ \mathcal{I}^{(\text{f-i})}_{i_1} $ that contain up to $ \ell $ losses it is enough to set a maximum distance of $ 3(\ell-1) $ from $ i_1 $. For $ \ell=3 $, the patch $ \boldsymbol{P} $ contains the qubits that are at a distance $ 6 $ or less from $ i_1 $. 

$ 2. $ Initialize an empty list $ \mathcal{I}  $ that will contain the instances from the patch, the number of edges that they erase and the associated energies.

$ 3. $ For every instance $ \{i,i'\}\subset\boldsymbol{P} $ with two different losses one has to compute $ R_{\{i,i'\}} $ from Eq.~(\ref{correction_average}) by averaging the number of edges erased over all possible corrections. Then, with the obtained $ R_{\{i,i'\}} $, one has to compute the energy of the instance $ \{i,i'\}$ that from Eq.~(\ref{E(R)}) takes the form:
\begin{equation}
	E_{\{i,i'\}}=R_{\{i,i'\}}-2R_1.
\end{equation}
Recall that for all instances $ \{i\} $ with only one loss, $ R_{\{i\}}=R_1 $ as explained in Section \ref{corrections}. Append the element $ \left[\{i,i'\},R_{\{i,i'\}},E_{\{i,i'\}}\right] $ to the list $ \mathcal{I}  $.

$ 4. $ For every instance $ \{i_1,i,i'\}\subset\boldsymbol{P} $ with three different losses (one of them the central loss $ i_1 $) one has to compute $ R_{\{i_1,i,i'\}} $ from Eq.~(\ref{correction_average}), then compute the energy of the instance from Eq.~(\ref{E(R)}), that takes the form:
\begin{equation}
E_{\{i_1,i,i'\}}=R_{\{i_1,i,i'\}}-R_{\{i_1,i\}}-R_{\{i_1,i'\}}-R_{\{i,i'\}}+3R_1 
\end{equation}
where we used again that for all instances $ \{i\} $ with only one loss, $ R_{\{i\}}=R_1 $. Note that the values of $ R_{\{i_1,i\}}$, $R_{\{i_1,i'\}}$, $R_{\{i,i'\}} $ are stored in $ \mathcal{I} $ for every $ i,i'\in\boldsymbol{P} $. Append the element $ \left[\{i_1,i,i'\},R_{\{i,i'\}},E_{\{i,i'\}}\right] $ to the list $ \mathcal{I}  $.

Finally, from the list $ \mathcal{I}  $, extract only those instances that contain the central loss $ i_1 $ and have non-zero energy. These constitute the set $ \mathcal{I}^{(\text{f-i})}_{i_1} $ that can be used to compute the coefficients $ \alpha_2 $ and $ \alpha_{3} $ with Eq.~(\ref{coefficients2}).


\section{Fundamental threshold for qubit loss}\label{sec_algebraic}
In this section we describe the algebraic technique employed to determine the existence of well-defined logical operators that do not have support on the set of removed qubits. This technique, which can be used to compute the fundamental qubit loss threshold $ p_f $,  determines efficiently if there exists a subset $ \mathcal{G} $ of generators such that the logical operator $ \tilde{l}_q^\sigma $ in Eq.~(\ref{equivalence}) does not have support on the set of removed qubits $ \boldsymbol{r} $ by mapping this problem to a system of linear binary equations. Furthermore, we prove the following statement: \textit{given a set of removed qubits $ \boldsymbol{r} $, the logical information is protected if and only if $ \boldsymbol{r} $ does not contain the support of any logical operator.}


\subsection{Algebraic technique}
Here we map the problem of finding $ \mathcal{G} $ to a system of linear binary equations. Without loss of generality we can choose the logical operator $ l_q^\sigma $ in Eq.~(\ref{equivalence}) as composed of Pauli operators of just one type $ \sigma $, like in Eq.~(\ref{logical}), where $ \boldsymbol{s}_q^\sigma $ is the set of qubits where $ l_q^\sigma $ has support. When a logical operator $ l_q^\sigma $ composed by Pauli operators of just one type $ \sigma $ is multiplied by generators of another type $ \sigma'\neq\sigma $, the support $ \boldsymbol{s}_q^\sigma $ of the new operator $ \tilde{l}_q^\sigma $ contains the support of $ l_q^\sigma $: $ \tilde{\boldsymbol{s}}_q^\sigma\supset\boldsymbol{s}_q^\sigma $, so if a removed qubit is in $ \boldsymbol{s}_q^\sigma $ it will also be in $ \tilde{\boldsymbol{s}}_q^\sigma $ and the multiplication with generators of other type $ \sigma' $ will be ineffective. 

 

As a consequence, we can restrict the subsets $ \mathcal{G} $ that multiply $ l_q^\sigma $ in Eq.~(\ref{equivalence}) to those subsets that only contain generators of the same type $ \sigma $. If the subset of faces where the generators of $ \mathcal{G} $ are defined is $ \mathcal{X} $, the support of $ \tilde{l}_q^\sigma $ is then given by:
\begin{equation}	\label{equivalence_sets}
\tilde{\boldsymbol{s}}_q^\sigma=\boldsymbol{s}_q^\sigma\bigoplus_{\boldsymbol{f}\in\mathcal{X}}\boldsymbol{f}
\end{equation}
where the symbol $ \oplus $ indicates the symmetric difference between sets: $ \boldsymbol{a}\oplus \boldsymbol{b}=(\boldsymbol{a}\cup \boldsymbol{b})\setminus(\boldsymbol{a}\cap \boldsymbol{b}) $. The symmetric difference comes from the fact that $ \sigma^n=\sigma $ for odd $ n $ and $ \sigma^n=I $ (the identity operator) for even $ n $. For simplicity, from now on we drop the indices $ q,\sigma $.

Given a set of removed qubits $ \boldsymbol{r} $, a logical operator $ \tilde{l}_q^\sigma $, defined on the string $ \tilde{\boldsymbol{s}} $, has non-empty support on $ \boldsymbol{r} $ if $ \tilde{\boldsymbol{s}} $ intersects $ \boldsymbol{r} $, i.e., if $ \boldsymbol{r}\cap\tilde{\boldsymbol{s}}\neq\emptyset $. Therefore, the logical information still exists if there is a subset of faces $ \mathcal{X} $ for which:
\begin{equation}\label{sets_system}
	\boldsymbol{r}\cap\left( \boldsymbol{s}\bigoplus_{\boldsymbol{f}\in\mathcal{X}}\boldsymbol{f}\right) =\emptyset	.
\end{equation} 

In order to map Eq.~(\ref{sets_system}) to a system of linear equations let us first define the binary vectors and matrices that represent the sets appearing in the equation. Recall that $ N $ is the number of qubits and $ F $ the number of faces. Then:
\begin{itemize}

	\item The set of all faces is represented by a $ N\times F $ matrix $ \mathbb{F} $ whose elements are $ \mathbb{F}_{i\boldsymbol{f}}=1$ if the qubit $ i $ is in the face $ \boldsymbol{f} $ and $0$  otherwise.
	\item A string $ \boldsymbol{s} $ is represented by a $ N\times 1 $ column vector $ \mathbb{s} $ whose elements are $ \mathbb{s}_{i}=1$ if the qubit $ i $ is in $ \boldsymbol{s} $ and $0$  otherwise.
	\item The subset $ \mathcal{X} $ of faces is represented by a $ F\times 1 $ column matrix $ x $ whose elements are $ x_{\boldsymbol{f}}=1 $ if the face $ \boldsymbol{f} $ is in $ \mathcal{X} $ and  $0$  otherwise.

\end{itemize}

The symmetric difference between sets is mapped to the summation modulo 2 of binary vectors and matrices. Then, Eq.~(\ref{equivalence_sets}) is mapped to the following binary matrix operations:
\begin{equation}
	\tilde{\mathbb{s}} = \mathbb{s} + \mathbb{F}x
\end{equation}
where $ \mathbb{F}x $ is the usual matrix product performed modulo 2.

The intersection between sets is mapped to the element-wise product $ \mathbb{r}\circ\tilde{\mathbb{s}} $ of binary vectors, i.e., another $ N\times 1 $ column vector where the $ i $-th element is the product $ \mathbb{r}_i\tilde{\mathbb{s}}_{i} $. Then, Eq.~(\ref{sets_system}) is mapped to
\begin{equation}
	\mathbb{r}\circ\left( \mathbb{s} + \mathbb{F}x\right) =0	,
\end{equation}
which can be written in the standard form of a system of linear equations as:
\begin{equation}\label{system}
	\left( \mathbb{r}\circ\mathbb{F}\right)x = \mathbb{r}\circ\mathbb{s}	.
\end{equation}
Here $ \mathbb{r}\circ\mathbb{F} $ is a $ N\times F $ matrix whose elements are the product $ \left[\mathbb{r}\circ\mathbb{F}\right]_{i\boldsymbol{f}}=\mathbb{r}_i\mathbb{F}_{i\boldsymbol{f}} $.

Finally, the search of a logical operator without support on the removed qubits is equivalent to finding a solution $ x $ of the linear system in Eq.~(\ref{system}). This system can be efficiently solved by Gauss elimination, in a time that scales as $ \sim N^3 $ or better.

\subsection{Necessary and sufficient condition for the existence of the logical information}
Here we prove that \textit{given a set of removed qubits  $ \boldsymbol{r} $, there exists a logical operator for every class $\{q,\sigma\}$ without support on the removed qubits if and only if $ \boldsymbol{r} $ does not contain the support of a logical operator}. We use the notation defined after Eq.~(\ref{sets_system}).

Let us start by assuming that $ \boldsymbol{r} $ includes the support of a logical operator $ l_q^\sigma $ and prove that all the logical operators $ l_q^{\sigma'} $ of other type $ \sigma'\neq \sigma $ have non-empty support on $ \boldsymbol{r} $. The logical operator $  l_q^{\sigma} $ anticommutes with all logical operators $ l_q^{\sigma'} $ of the class $ \{q,\sigma'\} $. Consequently the support of $  l_q^{\sigma} $ and the support of every logical operator $ l_q^{\sigma'} $ have some qubits in common. As a consequence, all logical operators $ l_q^{\sigma'} $ have non-empty support on the set of removed qubits, and therefore, the class $ \{q,\sigma'\} $ is not well defined.

Now we assume that the logical information does no longer exist, i.e., the system of Eq.~(\ref{system}) does not have a solution, and prove that the set of removed qubits represented by $ \mathbb{r} $ includes a logical operator. If the system has no solution, the rank of the augmented matrix is bigger than the rank of the matrix $ \mathbb{r}\circ\mathbb{F} $:
\begin{equation}\label{rank_inequality}
	\text{rank}\left( \mathbb{r}\circ\left[ \mathbb{F}\;\;\mathbb{s}\right] \right) > \text{rank}\left( \mathbb{r}\circ\mathbb{F} \right)	.
\end{equation}

By the rank-nullity theorem, the rank of any matrix $ A $ is the number of rows $ m $ minus the number of linearly independent column vectors $ \mathbb{v} $ that cancel it from the left: $ \mathbb{v}^TA=0 $. From Eq.~(\ref{rank_inequality}) this means that the matrix $ \mathbb{r}\circ\mathbb{F} $ has at least one more vector $ \mathbb{v} $ that cancels it from the left than the matrix $ \mathbb{r}\circ\left[ \mathbb{F}\;\;\mathbb{s}\right] $. Note that every vector that cancels $ \left[ \mathbb{F}\;\;\mathbb{s}\right] $ from the left also cancels $ \mathbb{F} $ from the left. Then, this vector satisfies that:
\begin{eqnarray}
	\mathbb{v}^T \left( \mathbb{r}\circ\mathbb{F} \right) = 0 \\
	\mathbb{v}^T\left( \mathbb{r}\circ\left[ \mathbb{F}\;\;\mathbb{s}\right] \right)\neq 0
\end{eqnarray}
or equivalently:
\begin{eqnarray}
	\mathbb{v}^T \left( \mathbb{r}\circ\mathbb{F} \right) = 0 \\
	\mathbb{v}^T\left( \mathbb{r}\circ\mathbb{s} \right)\neq 0
\end{eqnarray}
By using the commutation of the element-wise product $ \circ $ with the usual matrix product, we get that:
\begin{eqnarray}
\left( \mathbb{v} \circ\mathbb{r}\right)^T\mathbb{F}  = 0 \\
\left(\mathbb{v}\circ \mathbb{r}\right)^T\mathbb{s} \neq 0
\end{eqnarray}
which means that the vector $ \mathbb{v} \circ\mathbb{r} $ has an even number of qubits in common with the support of all generators represented by $ \mathbb{F} $, but an odd number in common with the support of the logical operator $ l_q^\sigma $ represented by $ \mathbb{s} $. The only possibility is that $ \mathbb{v} \circ\mathbb{r} $ is the support of a logical operator $ l_q^{\sigma'} $ of the class $ \{q,\sigma'\} $.

Given that if $ (\mathbb{v} \circ\mathbb{r})_i=1 $, then $ \mathbb{r}_i=1 $, the column vector $ \mathbb{r} $ represents a set of qubits $ \boldsymbol{r} $ that contains the support of the logical operator $ l_q^{\sigma'} $. Hence, we prove the statement in both logical directions.


\section{Conclusions and outlook} \label{conclusions}

In this work we have explored a connection between statistical mechanics and QEC arising from the study of qubit loss in the topological color code. Here the problem of determining the robustness of the code to qubit loss is mapped to a novel classical percolation problem on coupled lattices as recently proposed in \cite{Vodola2018}. By exploring this connection we have determined analytically the tolerance of the color code to qubit loss.

The main goal of this paper is to obtain analytically the critical qubit loss rate $ p_c $ below which the logical information in the color code is still protected. We have shown that $ p_c $ is related to the bond-percolation threshold $ r_c $ of the shrunk lattices of the color code through the equation $ r(p_c)=r_c $, where $ r(p) $ is the average fraction of edges erased at a qubit loss rate $ p $. We have developed a technique to systematically obtain the expansion coefficients of $ r(p) $, and we have presented an algorithm to calculate the values of these coefficients. We have computed the first three of these coefficients and found agreement with the numerical estimations.

Moreover, the fundamental loss threshold $ p_f $ of the three regular geometries of the color code has been computed numerically. Our results confirm the high robustness to qubit loss of the color code together with the protocol to correct qubit losses \cite{Vodola2018}, which is of practical relevance for actual and future quantum processors. Furthermore, in this paper we have proven that the logical information still exists after correcting the qubit losses if and only if the set of lost and sacrificed qubits together does not include the support of a logical operator.

Our work establishes the theoretical framework that might serve as a basis for future extensions of the protocol to correct losses. For example, the sacrificed qubits could be selected following global criteria that take into account the positions of all losses. Other extensions of the protocol could involve addressing more complex error models, e.g. taking into account possible (spatial) correlations between loss events, the imperfect identification of their positions, and the combined presence of qubit loss, computational, and measurement errors.

\begin{acknowledgments}
	We acknowledge support by U.S. A.R.O. through Grant No. W911NF-14-1-010. The research is also based upon work supported by the Office of the Director of National Intelligence (ODNI), Intelligence Advanced Research Projects Activity (IARPA), via the U.S. Army Research Office Grant No. W911NF-16-1-0070. The views and conclusions contained herein are those of the authors and should not be interpreted as necessarily representing the official policies or endorsements, either expressed or implied, of the ODNI, IARPA, or the U.S. Government. The U.S. Government is authorized to reproduce and distribute reprints for Governmental purposes notwithstanding any copyright annotation thereon. Any opinions, findings, and conclusions or recommendations expressed in this material are those of the author(s) and do not necessarily reflect the view of the U.S. Army Research Office. We acknowledge the resources and support of High Performance Computing Wales, where all the simulations were performed.
\end{acknowledgments}

\appendix

\section{Proof of Eq.~(\ref{E(R)})}\label{proof_E(R)}
In this Appendix we prove that the energy $ E_{\boldsymbol{i}} $ of a loss instance $ \boldsymbol{i} $ can be expressed in terms of the average number of edges $ R_{\boldsymbol{j}} $ as expressed in Eq.~(\ref{E(R)}). 

Let us rewrite Eqs.~\eqref{R(E)} and \eqref{E(R)} by using a delta function that equals $ 1 $ if $ \boldsymbol{j}\subset\boldsymbol{i} $ and zero otherwise:
\begin{eqnarray}
	R_{\boldsymbol{j}} &&= \sum_{\boldsymbol{k}\in\mathcal{I}}E_{\boldsymbol{k}} \delta_{\boldsymbol{k}\subset\boldsymbol{j}}\\
	E_{\boldsymbol{i}} &&= (-1)^{|\boldsymbol{i}|}\sum_{\boldsymbol{j}\in\mathcal{I}}(-1)^{|\boldsymbol{j}|}R_{\boldsymbol{j}} \delta_{\boldsymbol{j}\subset\boldsymbol{i}}	.
\end{eqnarray}
Here $ \mathcal{I} $ is the set of all loss instances. Substituting the first equation into the second one yields:
\begin{equation}
	E_{\boldsymbol{i}} = (-1)^{|\boldsymbol{i}|}\sum_{\boldsymbol{k}\in\mathcal{I}}E_{\boldsymbol{k}}\sum_{\boldsymbol{j}\in\mathcal{I}}(-1)^{|\boldsymbol{j}|} \delta_{\boldsymbol{k}\subset\boldsymbol{j}}\delta_{\boldsymbol{j}\subset\boldsymbol{i}}	.
\end{equation}

Instead of summing over $ \boldsymbol{j} $ we sum over the set difference $ \boldsymbol{t}=\boldsymbol{j}\setminus \boldsymbol{k} $, that contains all the subsets of $ \boldsymbol{i}\setminus \boldsymbol{k} $. Then, we have that:
\begin{equation}
\sum_{\boldsymbol{j}\in\mathcal{I}}(-1)^{|\boldsymbol{j}|} \delta_{\boldsymbol{k}\subset \boldsymbol{j}}\delta_{\boldsymbol{j}\subset \boldsymbol{i}} = \delta_{\boldsymbol{k}\subset \boldsymbol{i}}\sum_{\boldsymbol{t}\subset \boldsymbol{i}\setminus \boldsymbol{k}}(-1)^{|\boldsymbol{t}|+|\boldsymbol{k}|}	
\end{equation}
where $ \delta_{\boldsymbol{k}\subset \boldsymbol{i}} $ indicates that all the terms vanish if $ \boldsymbol{k}\not\subset\boldsymbol{i} $. The sum over $ \boldsymbol{t} $ equals zero unless $ |\boldsymbol{t}|=0 $, thus the number of elements of the sets $\boldsymbol{k}$ and $\boldsymbol{i}$ needs to be equal, i.e. $ |\boldsymbol{k}|=|\boldsymbol{i}| $:
\begin{equation}
\sum_{\boldsymbol{t}\subset \boldsymbol{i}\setminus \boldsymbol{k}}(-1)^{|\boldsymbol{t}|+|\boldsymbol{k}|} = (-1)^{|\boldsymbol{k}|}\delta_{|\boldsymbol{k}|=|\boldsymbol{i}|}	.
\end{equation}
Then, the sum over $ \boldsymbol{j} $ is reduced to a sign and two deltas:
\begin{equation}
E_{\boldsymbol{i}} = (-1)^{|\boldsymbol{i}|}\sum_{\boldsymbol{k}\in\mathcal{I}}E_{\boldsymbol{k}}(-1)^{|\boldsymbol{k}|}\delta_{\boldsymbol{k}\subset \boldsymbol{i}}\delta_{|\boldsymbol{k}|=|\boldsymbol{i}|}	.
\end{equation}
The condition imposed by the two deltas is satisfied if the sets  $\boldsymbol{k}$ and $\boldsymbol{i} $  are equal so the only term surviving in the sum over $ \boldsymbol{k} $ is $ \boldsymbol{k}=\boldsymbol{i} $. Hence the proof of Eq.~(\ref{E(R)}).

\section{Proof of Eq.~(\ref{coefficients})}\label{proof_expansion}
In this Appendix we prove that the $ \ell $-th coefficient $ \alpha_{\ell} $ in the expansion of the average fraction of edges erased $ r(p) $ in powers of $ p $ is given by the sum of energies $ E_{\boldsymbol{i}} $ of loss instances $ \boldsymbol{i} $ that contain $ \ell $ losses.

By substituting the number of edges erased $ R_{\boldsymbol{i}} $ in Eq.~(\ref{fraction_edges_erased}) by its expression in terms of energies in Eq.~(\ref{R(E)}) one gets that the average fraction of edges erased is:
\begin{equation}
	r(p)=e^{-1}\sum_{\boldsymbol{i}\in\mathcal{I}}p^{|\boldsymbol{i}|}(1-p)^{N-|\boldsymbol{i}|}\sum_{\boldsymbol{j}\subset\boldsymbol{i}}E_{\boldsymbol{j}}	.
\end{equation}
The condition in the second sum can be dropped by introducing a delta function $ \delta_{\boldsymbol{j}\subset \boldsymbol{i}} $ that equals 1 if $ \boldsymbol{j}\subset\boldsymbol{i} $ and 0 otherwise:
\begin{equation}
	r(p)=e^{-1}\sum_{\boldsymbol{j}\in\mathcal{I}}E_{\boldsymbol{j}}\sum_{\boldsymbol{i}\in\mathcal{I}}p^{|\boldsymbol{i}|}(1-p)^{N-|\boldsymbol{i}|}\delta_{\boldsymbol{j}\subset \boldsymbol{i}}	.
\end{equation}
For a fixed $ \boldsymbol{j} $ the instances $ \boldsymbol{i} $ for which the delta does not vanish are of the form $ \boldsymbol{i}=\boldsymbol{j}\cup\boldsymbol{k} $ where $ \boldsymbol{k} $ is a subset of the rest of qubits $ \boldsymbol{k}\subset\mathcal{V}\setminus\boldsymbol{j} $. Here $ \mathcal{V} $ is the set of all qubits. Then $ |\boldsymbol{i}|=|\boldsymbol{j}|+|\boldsymbol{k}| $ and the sum on $ \boldsymbol{i} $ can be substituted by a sum over $ \boldsymbol{k} $:
\begin{equation}
	r(p)=e^{-1}\sum_{\boldsymbol{j}\in\mathcal{I}}E_{\boldsymbol{j}}p^{|\boldsymbol{j}|}\sum_{\boldsymbol{k}\in\mathcal{V}\setminus\boldsymbol{j}}p^{|\boldsymbol{k}|}(1-p)^{(N-|\boldsymbol{j}|)-|\boldsymbol{k}|}	.
\end{equation}
The second sum equals one because it is a sum of the probabilities of every loss instance constrained to the qubits in $ \mathcal{V}\setminus\boldsymbol{j} $. This finalizes the proof of Eq.~(\ref{coefficients}).

\section{Separable instances have zero energy}\label{interacting}
In this Appendix we prove that the energy for a separable instance $ \boldsymbol{i} $ the energy $ E_{\boldsymbol{i}} $ vanishes. If two disjoint parts $ \boldsymbol{i}^{(A)},\,\boldsymbol{i}^{(B)} $ of an instance $ \boldsymbol{i}=\boldsymbol{i}^{(A)}\cup\boldsymbol{i}^{(B)} $ are far enough from each other, the number of edges erased is the sum of the edges erased by the two parts: $ R_{\boldsymbol{i}}=R_{\boldsymbol{i}^{(A)}}+R_{\boldsymbol{i}^{(B)}} $. This is defined as a separable instance.

In this situation, every loss in $ \boldsymbol{i}^{(A)} $ is far from every loss in $ \boldsymbol{i}^{(B)} $, so every subset $ \boldsymbol{j}\subset\boldsymbol{i} $ that contains some losses from $ \boldsymbol{i}^{(A)} $ and some losses from $ \boldsymbol{i}^{(B)} $:
\begin{equation}	\label{subinstance}
	\boldsymbol{j}\cap\boldsymbol{i}^{(A)}\neq\emptyset\;\;\;,\;\;\;\boldsymbol{j}\cap\boldsymbol{i}^{(B)}\neq\emptyset
\end{equation}
is also a separable instance:
\begin{equation}
	R_{\boldsymbol{j}}=R_{\boldsymbol{j}\cap\boldsymbol{i}^{(A)}}+R_{\boldsymbol{j}\cap\boldsymbol{i}^{(B)}}	.
\end{equation}

In particular, for the subsets $ \{j_1,j_2\} $ with just two losses, $ R_{\{j_1,j_2\}}=R_{\{j_1\}}+R_{\{j_2\}} $. So from Eq.~(\ref{E(R)}) we get that the energy of these subsets vanishes $ E_{\{j_1,j_2\}}=0 $.

For separable subsets $ \{j_1,j_2,j_3\} $ containing three losses $ R_{\{j_1,j_2,j_3\}}=R_{\{j_1\}}+R_{\{j_2,j_3\}} $. These subsets contain two subsets, $ \{j_1,j_2\},\,\{j_1,j_3\} $ whose energy vanishes. Then, using Eq.~(\ref{E(R)}) and canceling the vanishing energies at both two sides we have that the left and the right side of the previous equation are
\begin{eqnarray}
	&&\begin{split}
	R_{\{j_1,j_2,j_3\}}&=E_{\{j_1,j_2,j_3\}}+E_{\{j_2,j_3\}} \\
	&+E_{\{j_1\}}+E_{\{j_2\}}+E_{\{j_3\}},
	\end{split} \\
	&&\begin{split}
	R_{\{j_1\}}+R_{\{j_2,j_3\}} &= E_{\{j_1\}} + E_{\{j_2,j_3\}} \\
	&+E_{\{j_2\}}+E_{\{j_3\}},
	\end{split}
\end{eqnarray}
respectively. This results in a vanishing energy $ E_{\{j_1,j_2,j_3\}}=0 $.

Applying this derivation iteratively from subsets $ \boldsymbol{j}\subset\boldsymbol{i} $ of a separable instance $ \boldsymbol{i} $ we obtain that all energies $ E_{\boldsymbol{j}}=0 $ vanish. In particular, for the last iteration, when $ \boldsymbol{j}=\boldsymbol{i} $, the energy of $ \boldsymbol{i} $ vanishes $ E_{\boldsymbol{i}}=0 $, proving the initial statement.

\bibliography{library}{}

\begin{thebibliography}{56}%
\makeatletter
\providecommand \@ifxundefined [1]{%
 \@ifx{#1\undefined}
}%
\providecommand \@ifnum [1]{%
 \ifnum #1\expandafter \@firstoftwo
 \else \expandafter \@secondoftwo
 \fi
}%
\providecommand \@ifx [1]{%
 \ifx #1\expandafter \@firstoftwo
 \else \expandafter \@secondoftwo
 \fi
}%
\providecommand \natexlab [1]{#1}%
\providecommand \enquote  [1]{``#1''}%
\providecommand \bibnamefont  [1]{#1}%
\providecommand \bibfnamefont [1]{#1}%
\providecommand \citenamefont [1]{#1}%
\providecommand \href@noop [0]{\@secondoftwo}%
\providecommand \href [0]{\begingroup \@sanitize@url \@href}%
\providecommand \@href[1]{\@@startlink{#1}\@@href}%
\providecommand \@@href[1]{\endgroup#1\@@endlink}%
\providecommand \@sanitize@url [0]{\catcode `\\12\catcode `\$12\catcode
  `\&12\catcode `\#12\catcode `\^12\catcode `\_12\catcode `\%12\relax}%
\providecommand \@@startlink[1]{}%
\providecommand \@@endlink[0]{}%
\providecommand \url  [0]{\begingroup\@sanitize@url \@url }%
\providecommand \@url [1]{\endgroup\@href {#1}{\urlprefix }}%
\providecommand \urlprefix  [0]{URL }%
\providecommand \Eprint [0]{\href }%
\providecommand \doibase [0]{http://dx.doi.org/}%
\providecommand \selectlanguage [0]{\@gobble}%
\providecommand \bibinfo  [0]{\@secondoftwo}%
\providecommand \bibfield  [0]{\@secondoftwo}%
\providecommand \translation [1]{[#1]}%
\providecommand \BibitemOpen [0]{}%
\providecommand \bibitemStop [0]{}%
\providecommand \bibitemNoStop [0]{.\EOS\space}%
\providecommand \EOS [0]{\spacefactor3000\relax}%
\providecommand \BibitemShut  [1]{\csname bibitem#1\endcsname}%
\let\auto@bib@innerbib\@empty
\bibitem [{\citenamefont {Ladd}\ \emph {et~al.}(2010)\citenamefont {Ladd},
  \citenamefont {Jelezko}, \citenamefont {Laflamme}, \citenamefont {Nakamura},
  \citenamefont {Monroe},\ and\ \citenamefont {O'Brien}}]{Ladd2010}%
  \BibitemOpen
  \bibfield  {author} {\bibinfo {author} {\bibfnamefont {T.~D.}\ \bibnamefont
  {Ladd}}, \bibinfo {author} {\bibfnamefont {F.}~\bibnamefont {Jelezko}},
  \bibinfo {author} {\bibfnamefont {R.}~\bibnamefont {Laflamme}}, \bibinfo
  {author} {\bibfnamefont {Y.}~\bibnamefont {Nakamura}}, \bibinfo {author}
  {\bibfnamefont {C.}~\bibnamefont {Monroe}}, \ and\ \bibinfo {author}
  {\bibfnamefont {J.~L.}\ \bibnamefont {O'Brien}},\ }\href {\doibase
  10.1038/nature08812} {\bibfield  {journal} {\bibinfo  {journal} {Nature}\
  }\textbf {\bibinfo {volume} {464}},\ \bibinfo {pages} {45} (\bibinfo {year}
  {2010})}\BibitemShut {NoStop}%
\bibitem [{\citenamefont {Lewenstein}\ \emph {et~al.}(2007)\citenamefont
  {Lewenstein}, \citenamefont {Sanpera}, \citenamefont {Ahufinger},
  \citenamefont {Damski}, \citenamefont {Sen(De)},\ and\ \citenamefont
  {Sen}}]{Lewenstein2007}%
  \BibitemOpen
  \bibfield  {author} {\bibinfo {author} {\bibfnamefont {M.}~\bibnamefont
  {Lewenstein}}, \bibinfo {author} {\bibfnamefont {A.}~\bibnamefont {Sanpera}},
  \bibinfo {author} {\bibfnamefont {V.}~\bibnamefont {Ahufinger}}, \bibinfo
  {author} {\bibfnamefont {B.}~\bibnamefont {Damski}}, \bibinfo {author}
  {\bibfnamefont {A.}~\bibnamefont {Sen(De)}}, \ and\ \bibinfo {author}
  {\bibfnamefont {U.}~\bibnamefont {Sen}},\ }\href {\doibase
  10.1080/00018730701223200} {\bibfield  {journal} {\bibinfo  {journal} {Adv.
  Phys.}\ }\textbf {\bibinfo {volume} {56}},\ \bibinfo {pages} {243} (\bibinfo
  {year} {2007})}\BibitemShut {NoStop}%
\bibitem [{\citenamefont {Amico}\ \emph {et~al.}(2008)\citenamefont {Amico},
  \citenamefont {Fazio}, \citenamefont {Osterloh},\ and\ \citenamefont
  {Vedral}}]{Amico2008}%
  \BibitemOpen
  \bibfield  {author} {\bibinfo {author} {\bibfnamefont {L.}~\bibnamefont
  {Amico}}, \bibinfo {author} {\bibfnamefont {R.}~\bibnamefont {Fazio}},
  \bibinfo {author} {\bibfnamefont {A.}~\bibnamefont {Osterloh}}, \ and\
  \bibinfo {author} {\bibfnamefont {V.}~\bibnamefont {Vedral}},\ }\href
  {\doibase 10.1103/RevModPhys.80.517} {\bibfield  {journal} {\bibinfo
  {journal} {Rev. Mod. Phys.}\ }\textbf {\bibinfo {volume} {80}},\ \bibinfo
  {pages} {517} (\bibinfo {year} {2008})}\BibitemShut {NoStop}%
\bibitem [{\citenamefont {Nielsen}\ and\ \citenamefont
  {Chuang}(2000)}]{Nielsen2000}%
  \BibitemOpen
  \bibfield  {author} {\bibinfo {author} {\bibfnamefont {M.~A.}\ \bibnamefont
  {Nielsen}}\ and\ \bibinfo {author} {\bibfnamefont {I.~L.}\ \bibnamefont
  {Chuang}},\ }\href@noop {} {\emph {\bibinfo {title} {Quantum Computation and
  Quantum Information}}}\ (\bibinfo  {publisher} {Cambridge University Press},\
  \bibinfo {year} {2000})\BibitemShut {NoStop}%
\bibitem [{\citenamefont {{G. De las Cuevas}}(2013)}]{Cuevas2013}%
  \BibitemOpen
  \bibfield  {author} {\bibinfo {author} {\bibnamefont {{G. De las Cuevas}}},\
  }\href {\doibase 10.1088/0953-4075/46/24/243001} {\bibfield  {journal}
  {\bibinfo  {journal} {J. Phys. B}\ }\textbf {\bibinfo {volume} {46}},\
  \bibinfo {pages} {243001} (\bibinfo {year} {2013})}\BibitemShut {NoStop}%
\bibitem [{\citenamefont {Chubb}\ and\ \citenamefont
  {Flammia}(2019)}]{Chubb2019}%
  \BibitemOpen
  \bibfield  {author} {\bibinfo {author} {\bibfnamefont {C.~T.}\ \bibnamefont
  {Chubb}}\ and\ \bibinfo {author} {\bibfnamefont {S.~T.}\ \bibnamefont
  {Flammia}},\ }\href@noop {} {\  (\bibinfo {year} {2019})},\ \Eprint
  {http://arxiv.org/abs/arXiv:1809.10704v2} {arXiv:1809.10704v2} \BibitemShut
  {NoStop}%
\bibitem [{\citenamefont {Zarei}\ and\ \citenamefont
  {Montakhab}(2019)}]{Zarei2019}%
  \BibitemOpen
  \bibfield  {author} {\bibinfo {author} {\bibfnamefont {M.~H.}\ \bibnamefont
  {Zarei}}\ and\ \bibinfo {author} {\bibfnamefont {A.}~\bibnamefont
  {Montakhab}},\ }\href {\doibase 10.1103/PhysRevA.99.052312} {\bibfield
  {journal} {\bibinfo  {journal} {Phys. Rev. A}\ }\textbf {\bibinfo {volume}
  {99}},\ \bibinfo {pages} {052312} (\bibinfo {year} {2019})}\BibitemShut
  {NoStop}%
\bibitem [{\citenamefont {Van~den Nest}\ \emph {et~al.}(2008)\citenamefont
  {Van~den Nest}, \citenamefont {D\"ur},\ and\ \citenamefont
  {Briegel}}]{Nest2008-04}%
  \BibitemOpen
  \bibfield  {author} {\bibinfo {author} {\bibfnamefont {M.}~\bibnamefont
  {Van~den Nest}}, \bibinfo {author} {\bibfnamefont {W.}~\bibnamefont {D\"ur}},
  \ and\ \bibinfo {author} {\bibfnamefont {H.~J.}\ \bibnamefont {Briegel}},\
  }\href {\doibase 10.1103/PhysRevLett.100.110501} {\bibfield  {journal}
  {\bibinfo  {journal} {Phys. Rev. Lett.}\ }\textbf {\bibinfo {volume} {100}},\
  \bibinfo {pages} {110501} (\bibinfo {year} {2008})}\BibitemShut {NoStop}%
\bibitem [{\citenamefont {{De las Cuevas}}\ \emph {et~al.}(2009)\citenamefont
  {{De las Cuevas}}, \citenamefont {D\"ur}, \citenamefont {Briegel},\ and\
  \citenamefont {Martin-Delgado}}]{Cuevas2009}%
  \BibitemOpen
  \bibfield  {author} {\bibinfo {author} {\bibfnamefont {G.}~\bibnamefont {{De
  las Cuevas}}}, \bibinfo {author} {\bibfnamefont {W.}~\bibnamefont {D\"ur}},
  \bibinfo {author} {\bibfnamefont {H.~J.}\ \bibnamefont {Briegel}}, \ and\
  \bibinfo {author} {\bibfnamefont {M.~A.}\ \bibnamefont {Martin-Delgado}},\
  }\href {\doibase 10.1103/PhysRevLett.102.230502} {\bibfield  {journal}
  {\bibinfo  {journal} {Phys. Rev. Lett.}\ }\textbf {\bibinfo {volume} {102}},\
  \bibinfo {pages} {230502} (\bibinfo {year} {2009})}\BibitemShut {NoStop}%
\bibitem [{\citenamefont {Xu}\ \emph {et~al.}(2011)\citenamefont {Xu},
  \citenamefont {{G. De las Cuevas}}, \citenamefont {D{\"u}r}, \citenamefont
  {Briegel},\ and\ \citenamefont {Martin-Delgado}}]{Xu2011}%
  \BibitemOpen
  \bibfield  {author} {\bibinfo {author} {\bibfnamefont {Y.}~\bibnamefont
  {Xu}}, \bibinfo {author} {\bibnamefont {{G. De las Cuevas}}}, \bibinfo
  {author} {\bibfnamefont {W.}~\bibnamefont {D{\"u}r}}, \bibinfo {author}
  {\bibfnamefont {H.~J.}\ \bibnamefont {Briegel}}, \ and\ \bibinfo {author}
  {\bibfnamefont {M.~A.}\ \bibnamefont {Martin-Delgado}},\ }\href {\doibase
  10.1088/1742-5468/2011/02/p02013} {\bibfield  {journal} {\bibinfo  {journal}
  {J. Stat. Mech-Theory E}\ }\textbf {\bibinfo {volume} {2011}},\ \bibinfo
  {pages} {P02013} (\bibinfo {year} {2011})}\BibitemShut {NoStop}%
\bibitem [{\citenamefont {{De las Cuevas}}\ and\ \citenamefont
  {Cubitt}(2016)}]{Cuevas2016}%
  \BibitemOpen
  \bibfield  {author} {\bibinfo {author} {\bibfnamefont {G.}~\bibnamefont {{De
  las Cuevas}}}\ and\ \bibinfo {author} {\bibfnamefont {T.~S.}\ \bibnamefont
  {Cubitt}},\ }\href {\doibase 10.1126/science.aab3326} {\bibfield  {journal}
  {\bibinfo  {journal} {Science}\ }\textbf {\bibinfo {volume} {351}},\ \bibinfo
  {pages} {1180} (\bibinfo {year} {2016})}\BibitemShut {NoStop}%
\bibitem [{\citenamefont {den Nest}\ \emph {et~al.}(2007)\citenamefont {den
  Nest}, \citenamefont {D\"ur},\ and\ \citenamefont {Briegel}}]{Nest2007}%
  \BibitemOpen
  \bibfield  {author} {\bibinfo {author} {\bibfnamefont {M.~V.}\ \bibnamefont
  {den Nest}}, \bibinfo {author} {\bibfnamefont {W.}~\bibnamefont {D\"ur}}, \
  and\ \bibinfo {author} {\bibfnamefont {H.~J.}\ \bibnamefont {Briegel}},\
  }\href {\doibase 10.1103/PhysRevLett.98.117207} {\bibfield  {journal}
  {\bibinfo  {journal} {Phys. Rev. Lett.}\ }\textbf {\bibinfo {volume} {98}},\
  \bibinfo {pages} {117207} (\bibinfo {year} {2007})}\BibitemShut {NoStop}%
\bibitem [{\citenamefont {{G. De las Cuevas}}\ \emph
  {et~al.}(2011)\citenamefont {{G. De las Cuevas}}, \citenamefont {D{\"u}r},
  \citenamefont {den Nest},\ and\ \citenamefont {Martin-Delgado}}]{Cuevas2011}%
  \BibitemOpen
  \bibfield  {author} {\bibinfo {author} {\bibnamefont {{G. De las Cuevas}}},
  \bibinfo {author} {\bibfnamefont {W.}~\bibnamefont {D{\"u}r}}, \bibinfo
  {author} {\bibfnamefont {M.~V.}\ \bibnamefont {den Nest}}, \ and\ \bibinfo
  {author} {\bibfnamefont {M.~A.}\ \bibnamefont {Martin-Delgado}},\ }\href
  {\doibase 10.1088/1367-2630/13/9/093021} {\bibfield  {journal} {\bibinfo
  {journal} {New J. Phys.}\ }\textbf {\bibinfo {volume} {13}},\ \bibinfo
  {pages} {093021} (\bibinfo {year} {2011})}\BibitemShut {NoStop}%
\bibitem [{\citenamefont {Geraci}\ and\ \citenamefont
  {Lidar}(2008)}]{Geraci2008}%
  \BibitemOpen
  \bibfield  {author} {\bibinfo {author} {\bibfnamefont {J.}~\bibnamefont
  {Geraci}}\ and\ \bibinfo {author} {\bibfnamefont {D.~A.}\ \bibnamefont
  {Lidar}},\ }\href {\doibase 10.1007/s00220-008-0438-0} {\bibfield  {journal}
  {\bibinfo  {journal} {Commun. Math. Phys.}\ }\textbf {\bibinfo {volume}
  {279}},\ \bibinfo {pages} {735} (\bibinfo {year} {2008})}\BibitemShut
  {NoStop}%
\bibitem [{\citenamefont {Lidar}\ and\ \citenamefont
  {Biham}(1997)}]{Lidar1997}%
  \BibitemOpen
  \bibfield  {author} {\bibinfo {author} {\bibfnamefont {D.~A.}\ \bibnamefont
  {Lidar}}\ and\ \bibinfo {author} {\bibfnamefont {O.}~\bibnamefont {Biham}},\
  }\href {\doibase 10.1103/PhysRevE.56.3661} {\bibfield  {journal} {\bibinfo
  {journal} {Phys. Rev. E}\ }\textbf {\bibinfo {volume} {56}},\ \bibinfo
  {pages} {3661} (\bibinfo {year} {1997})}\BibitemShut {NoStop}%
\bibitem [{\citenamefont {Somma}\ \emph {et~al.}(2007)\citenamefont {Somma},
  \citenamefont {Batista},\ and\ \citenamefont {Ortiz}}]{Somma2007}%
  \BibitemOpen
  \bibfield  {author} {\bibinfo {author} {\bibfnamefont {R.~D.}\ \bibnamefont
  {Somma}}, \bibinfo {author} {\bibfnamefont {C.~D.}\ \bibnamefont {Batista}},
  \ and\ \bibinfo {author} {\bibfnamefont {G.}~\bibnamefont {Ortiz}},\ }\href
  {\doibase 10.1103/PhysRevLett.99.030603} {\bibfield  {journal} {\bibinfo
  {journal} {Phys. Rev. Lett.}\ }\textbf {\bibinfo {volume} {99}},\ \bibinfo
  {pages} {030603} (\bibinfo {year} {2007})}\BibitemShut {NoStop}%
\bibitem [{\citenamefont {Geraci}\ and\ \citenamefont
  {Lidar}(2010)}]{Geraci2010}%
  \BibitemOpen
  \bibfield  {author} {\bibinfo {author} {\bibfnamefont {J.}~\bibnamefont
  {Geraci}}\ and\ \bibinfo {author} {\bibfnamefont {D.~A.}\ \bibnamefont
  {Lidar}},\ }\href {\doibase 10.1088/1367-2630/12/7/075026} {\bibfield
  {journal} {\bibinfo  {journal} {New J. Phys.}\ }\textbf {\bibinfo {volume}
  {12}},\ \bibinfo {pages} {075026} (\bibinfo {year} {2010})}\BibitemShut
  {NoStop}%
\bibitem [{\citenamefont {Browne}(2014)}]{Browne2014}%
  \BibitemOpen
  \bibfield  {author} {\bibinfo {author} {\bibfnamefont {D.}~\bibnamefont
  {Browne}},\ }\href@noop {} {\  (\bibinfo {year} {2014})},\ \Eprint
  {http://arxiv.org/abs/arXiv:1401.2466v1} {arXiv:1401.2466v1} \BibitemShut
  {NoStop}%
\bibitem [{\citenamefont {Terhal}(2015)}]{Terhal2015}%
  \BibitemOpen
  \bibfield  {author} {\bibinfo {author} {\bibfnamefont {B.~M.}\ \bibnamefont
  {Terhal}},\ }\href {\doibase 10.1103/RevModPhys.87.307} {\bibfield  {journal}
  {\bibinfo  {journal} {Rev. Mod. Phys.}\ }\textbf {\bibinfo {volume} {87}},\
  \bibinfo {pages} {307} (\bibinfo {year} {2015})}\BibitemShut {NoStop}%
\bibitem [{\citenamefont {Jahromi}\ \emph {et~al.}(2013)\citenamefont
  {Jahromi}, \citenamefont {Masoudi}, \citenamefont {Kargarian},\ and\
  \citenamefont {Schmidt}}]{Jahromi2013}%
  \BibitemOpen
  \bibfield  {author} {\bibinfo {author} {\bibfnamefont {S.~S.}\ \bibnamefont
  {Jahromi}}, \bibinfo {author} {\bibfnamefont {S.~F.}\ \bibnamefont
  {Masoudi}}, \bibinfo {author} {\bibfnamefont {M.}~\bibnamefont {Kargarian}},
  \ and\ \bibinfo {author} {\bibfnamefont {K.~P.}\ \bibnamefont {Schmidt}},\
  }\href {\doibase 10.1103/PhysRevB.88.214411} {\bibfield  {journal} {\bibinfo
  {journal} {Phys. Rev. B}\ }\textbf {\bibinfo {volume} {88}},\ \bibinfo
  {pages} {214411} (\bibinfo {year} {2013})}\BibitemShut {NoStop}%
\bibitem [{\citenamefont {Zarei}\ and\ \citenamefont
  {Montakhab}(2018)}]{Zarei2018}%
  \BibitemOpen
  \bibfield  {author} {\bibinfo {author} {\bibfnamefont {M.~H.}\ \bibnamefont
  {Zarei}}\ and\ \bibinfo {author} {\bibfnamefont {A.}~\bibnamefont
  {Montakhab}},\ }\href {\doibase 10.1103/PhysRevA.98.012337} {\bibfield
  {journal} {\bibinfo  {journal} {Phys. Rev. A}\ }\textbf {\bibinfo {volume}
  {98}},\ \bibinfo {pages} {1} (\bibinfo {year} {2018})}\BibitemShut {NoStop}%
\bibitem [{\citenamefont {Kitaev}(2003)}]{Kitaev2003}%
  \BibitemOpen
  \bibfield  {author} {\bibinfo {author} {\bibfnamefont {A.}~\bibnamefont
  {Kitaev}},\ }\href {\doibase 10.1016/S0003-4916(02)00018-0} {\bibfield
  {journal} {\bibinfo  {journal} {Ann. Phys.}\ }\textbf {\bibinfo {volume}
  {303}},\ \bibinfo {pages} {2 } (\bibinfo {year} {2003})}\BibitemShut
  {NoStop}%
\bibitem [{\citenamefont {Bombin}\ and\ \citenamefont
  {Martin-Delgado}(2006)}]{Bombin2006}%
  \BibitemOpen
  \bibfield  {author} {\bibinfo {author} {\bibfnamefont {H.}~\bibnamefont
  {Bombin}}\ and\ \bibinfo {author} {\bibfnamefont {M.~A.}\ \bibnamefont
  {Martin-Delgado}},\ }\href {\doibase 10.1103/PhysRevLett.97.180501}
  {\bibfield  {journal} {\bibinfo  {journal} {Phys. Rev. Lett.}\ }\textbf
  {\bibinfo {volume} {97}},\ \bibinfo {pages} {180501} (\bibinfo {year}
  {2006})}\BibitemShut {NoStop}%
\bibitem [{\citenamefont {Bombin}\ and\ \citenamefont
  {Martin-Delgado}(2007)}]{Bombin2007}%
  \BibitemOpen
  \bibfield  {author} {\bibinfo {author} {\bibfnamefont {H.}~\bibnamefont
  {Bombin}}\ and\ \bibinfo {author} {\bibfnamefont {M.~A.}\ \bibnamefont
  {Martin-Delgado}},\ }\href {\doibase 10.1103/PhysRevLett.98.160502}
  {\bibfield  {journal} {\bibinfo  {journal} {Phys. Rev. Lett.}\ }\textbf
  {\bibinfo {volume} {98}},\ \bibinfo {pages} {160502} (\bibinfo {year}
  {2007})}\BibitemShut {NoStop}%
\bibitem [{\citenamefont {Dennis}\ \emph {et~al.}(2002)\citenamefont {Dennis},
  \citenamefont {Kitaev}, \citenamefont {Landahl},\ and\ \citenamefont
  {Preskill}}]{Dennis2002}%
  \BibitemOpen
  \bibfield  {author} {\bibinfo {author} {\bibfnamefont {E.}~\bibnamefont
  {Dennis}}, \bibinfo {author} {\bibfnamefont {A.}~\bibnamefont {Kitaev}},
  \bibinfo {author} {\bibfnamefont {A.}~\bibnamefont {Landahl}}, \ and\
  \bibinfo {author} {\bibfnamefont {J.}~\bibnamefont {Preskill}},\ }\href
  {\doibase 10.1063/1.1499754} {\bibfield  {journal} {\bibinfo  {journal} {J.
  Math. Phys.}\ }\textbf {\bibinfo {volume} {43}},\ \bibinfo {pages} {4452}
  (\bibinfo {year} {2002})}\BibitemShut {NoStop}%
\bibitem [{\citenamefont {Katzgraber}\ \emph {et~al.}(2009)\citenamefont
  {Katzgraber}, \citenamefont {Bombin},\ and\ \citenamefont
  {Martin-Delgado}}]{Katzgraber2009}%
  \BibitemOpen
  \bibfield  {author} {\bibinfo {author} {\bibfnamefont {H.~G.}\ \bibnamefont
  {Katzgraber}}, \bibinfo {author} {\bibfnamefont {H.}~\bibnamefont {Bombin}},
  \ and\ \bibinfo {author} {\bibfnamefont {M.~A.}\ \bibnamefont
  {Martin-Delgado}},\ }\href {\doibase 10.1103/PhysRevLett.103.090501}
  {\bibfield  {journal} {\bibinfo  {journal} {Phys. Rev. Lett.}\ }\textbf
  {\bibinfo {volume} {103}},\ \bibinfo {pages} {090501} (\bibinfo {year}
  {2009})}\BibitemShut {NoStop}%
\bibitem [{\citenamefont {Ohno}\ \emph {et~al.}(2004)\citenamefont {Ohno},
  \citenamefont {Arakawa}, \citenamefont {Ichinose},\ and\ \citenamefont
  {Matsui}}]{Ohno2004}%
  \BibitemOpen
  \bibfield  {author} {\bibinfo {author} {\bibfnamefont {T.}~\bibnamefont
  {Ohno}}, \bibinfo {author} {\bibfnamefont {G.}~\bibnamefont {Arakawa}},
  \bibinfo {author} {\bibfnamefont {I.}~\bibnamefont {Ichinose}}, \ and\
  \bibinfo {author} {\bibfnamefont {T.}~\bibnamefont {Matsui}},\ }\href
  {\doibase http://dx.doi.org/10.1016/j.nuclphysb.2004.07.003} {\bibfield
  {journal} {\bibinfo  {journal} {Nucl. Phys. B}\ }\textbf {\bibinfo {volume}
  {697}},\ \bibinfo {pages} {462} (\bibinfo {year} {2004})}\BibitemShut
  {NoStop}%
\bibitem [{\citenamefont {Andrist}\ \emph {et~al.}(2011)\citenamefont
  {Andrist}, \citenamefont {Katzgraber}, \citenamefont {Bombin},\ and\
  \citenamefont {Martin-Delgado}}]{Andrist2011}%
  \BibitemOpen
  \bibfield  {author} {\bibinfo {author} {\bibfnamefont {R.~S.}\ \bibnamefont
  {Andrist}}, \bibinfo {author} {\bibfnamefont {H.~G.}\ \bibnamefont
  {Katzgraber}}, \bibinfo {author} {\bibfnamefont {H.}~\bibnamefont {Bombin}},
  \ and\ \bibinfo {author} {\bibfnamefont {M.~A.}\ \bibnamefont
  {Martin-Delgado}},\ }\href {http://stacks.iop.org/1367-2630/13/i=8/a=083006}
  {\bibfield  {journal} {\bibinfo  {journal} {New J. Phys.}\ }\textbf {\bibinfo
  {volume} {13}},\ \bibinfo {pages} {83006} (\bibinfo {year}
  {2011})}\BibitemShut {NoStop}%
\bibitem [{\citenamefont {Bennett}\ \emph {et~al.}(1997)\citenamefont
  {Bennett}, \citenamefont {DiVincenzo},\ and\ \citenamefont
  {Smolin}}]{Bennett1997}%
  \BibitemOpen
  \bibfield  {author} {\bibinfo {author} {\bibfnamefont {C.~H.}\ \bibnamefont
  {Bennett}}, \bibinfo {author} {\bibfnamefont {D.~P.}\ \bibnamefont
  {DiVincenzo}}, \ and\ \bibinfo {author} {\bibfnamefont {J.~A.}\ \bibnamefont
  {Smolin}},\ }\href {\doibase 10.1103/PhysRevLett.78.3217} {\bibfield
  {journal} {\bibinfo  {journal} {Phys. Rev. Lett.}\ }\textbf {\bibinfo
  {volume} {78}},\ \bibinfo {pages} {3217} (\bibinfo {year}
  {1997})}\BibitemShut {NoStop}%
\bibitem [{\citenamefont {Almheiri}\ \emph {et~al.}(2015)\citenamefont
  {Almheiri}, \citenamefont {Dong},\ and\ \citenamefont
  {Harlow}}]{Almheiri2015}%
  \BibitemOpen
  \bibfield  {author} {\bibinfo {author} {\bibfnamefont {A.}~\bibnamefont
  {Almheiri}}, \bibinfo {author} {\bibfnamefont {X.}~\bibnamefont {Dong}}, \
  and\ \bibinfo {author} {\bibfnamefont {D.}~\bibnamefont {Harlow}},\ }\href
  {\doibase 10.1007/JHEP04(2015)163} {\bibfield  {journal} {\bibinfo  {journal}
  {J. High Energy Phys.}\ }\textbf {\bibinfo {volume} {2015}},\ \bibinfo
  {pages} {163} (\bibinfo {year} {2015})}\BibitemShut {NoStop}%
\bibitem [{\citenamefont {Pastawski}\ \emph {et~al.}(2015)\citenamefont
  {Pastawski}, \citenamefont {Yoshida}, \citenamefont {Harlow},\ and\
  \citenamefont {Preskill}}]{Pastawski2015}%
  \BibitemOpen
  \bibfield  {author} {\bibinfo {author} {\bibfnamefont {F.}~\bibnamefont
  {Pastawski}}, \bibinfo {author} {\bibfnamefont {B.}~\bibnamefont {Yoshida}},
  \bibinfo {author} {\bibfnamefont {D.}~\bibnamefont {Harlow}}, \ and\ \bibinfo
  {author} {\bibfnamefont {J.}~\bibnamefont {Preskill}},\ }\href {\doibase
  10.1007/JHEP06(2015)149} {\bibfield  {journal} {\bibinfo  {journal} {J. High
  Energy Phys.}\ }\textbf {\bibinfo {volume} {2015}},\ \bibinfo {pages} {149}
  (\bibinfo {year} {2015})}\BibitemShut {NoStop}%
\bibitem [{\citenamefont {Brown}\ \emph {et~al.}(2016)\citenamefont {Brown},
  \citenamefont {Kim},\ and\ \citenamefont {Monroe}}]{Brown2016}%
  \BibitemOpen
  \bibfield  {author} {\bibinfo {author} {\bibfnamefont {K.~R.}\ \bibnamefont
  {Brown}}, \bibinfo {author} {\bibfnamefont {J.}~\bibnamefont {Kim}}, \ and\
  \bibinfo {author} {\bibfnamefont {C.}~\bibnamefont {Monroe}},\ }\href
  {http://dx.doi.org/10.1038/npjqi.2016.34} {\bibfield  {journal} {\bibinfo
  {journal} {New J. Phys.}\ }\textbf {\bibinfo {volume} {2}},\ \bibinfo {pages}
  {16034} (\bibinfo {year} {2016})}\BibitemShut {NoStop}%
\bibitem [{\citenamefont {Barz}(2015)}]{Stefanie2015}%
  \BibitemOpen
  \bibfield  {author} {\bibinfo {author} {\bibfnamefont {S.}~\bibnamefont
  {Barz}},\ }\href {http://stacks.iop.org/0953-4075/48/i=8/a=083001} {\bibfield
   {journal} {\bibinfo  {journal} {J. Phys. B}\ }\textbf {\bibinfo {volume}
  {48}},\ \bibinfo {pages} {083001} (\bibinfo {year} {2015})}\BibitemShut
  {NoStop}%
\bibitem [{\citenamefont {Bloch}\ \emph {et~al.}(2008)\citenamefont {Bloch},
  \citenamefont {Dalibard},\ and\ \citenamefont {Zwerger}}]{Bloch2008}%
  \BibitemOpen
  \bibfield  {author} {\bibinfo {author} {\bibfnamefont {I.}~\bibnamefont
  {Bloch}}, \bibinfo {author} {\bibfnamefont {J.}~\bibnamefont {Dalibard}}, \
  and\ \bibinfo {author} {\bibfnamefont {W.}~\bibnamefont {Zwerger}},\ }\href
  {\doibase 10.1103/RevModPhys.80.885} {\bibfield  {journal} {\bibinfo
  {journal} {Rev. Mod. Phys.}\ }\textbf {\bibinfo {volume} {80}},\ \bibinfo
  {pages} {885} (\bibinfo {year} {2008})}\BibitemShut {NoStop}%
\bibitem [{\citenamefont {Clarke}\ and\ \citenamefont
  {Wilhelm}(2008)}]{Clarke2008}%
  \BibitemOpen
  \bibfield  {author} {\bibinfo {author} {\bibfnamefont {J.}~\bibnamefont
  {Clarke}}\ and\ \bibinfo {author} {\bibfnamefont {F.~K.}\ \bibnamefont
  {Wilhelm}},\ }\href {\doibase 10.1038/nature07128} {\bibfield  {journal}
  {\bibinfo  {journal} {Nature}\ }\textbf {\bibinfo {volume} {453}},\ \bibinfo
  {pages} {1031} (\bibinfo {year} {2008})}\BibitemShut {NoStop}%
\bibitem [{\citenamefont {Sherman}\ \emph {et~al.}(2013)\citenamefont
  {Sherman}, \citenamefont {Curtis}, \citenamefont {Szwer}, \citenamefont
  {Allcock}, \citenamefont {Imreh}, \citenamefont {Lucas},\ and\ \citenamefont
  {Steane}}]{Sherman2013}%
  \BibitemOpen
  \bibfield  {author} {\bibinfo {author} {\bibfnamefont {J.~A.}\ \bibnamefont
  {Sherman}}, \bibinfo {author} {\bibfnamefont {M.~J.}\ \bibnamefont {Curtis}},
  \bibinfo {author} {\bibfnamefont {D.~J.}\ \bibnamefont {Szwer}}, \bibinfo
  {author} {\bibfnamefont {D.~T.~C.}\ \bibnamefont {Allcock}}, \bibinfo
  {author} {\bibfnamefont {G.}~\bibnamefont {Imreh}}, \bibinfo {author}
  {\bibfnamefont {D.~M.}\ \bibnamefont {Lucas}}, \ and\ \bibinfo {author}
  {\bibfnamefont {A.~M.}\ \bibnamefont {Steane}},\ }\href {\doibase
  10.1103/PhysRevLett.111.180501} {\bibfield  {journal} {\bibinfo  {journal}
  {Phys. Rev. Lett.}\ }\textbf {\bibinfo {volume} {111}},\ \bibinfo {pages} {1}
  (\bibinfo {year} {2013})}\BibitemShut {NoStop}%
\bibitem [{\citenamefont {Ghosh}\ \emph {et~al.}(2013)\citenamefont {Ghosh},
  \citenamefont {Fowler}, \citenamefont {Martinis},\ and\ \citenamefont
  {Geller}}]{Ghosh2013}%
  \BibitemOpen
  \bibfield  {author} {\bibinfo {author} {\bibfnamefont {J.}~\bibnamefont
  {Ghosh}}, \bibinfo {author} {\bibfnamefont {A.~G.}\ \bibnamefont {Fowler}},
  \bibinfo {author} {\bibfnamefont {J.~M.}\ \bibnamefont {Martinis}}, \ and\
  \bibinfo {author} {\bibfnamefont {M.~R.}\ \bibnamefont {Geller}},\ }\href
  {\doibase 10.1103/PhysRevA.88.062329} {\bibfield  {journal} {\bibinfo
  {journal} {Phys. Rev. A}\ }\textbf {\bibinfo {volume} {88}},\ \bibinfo
  {pages} {062329} (\bibinfo {year} {2013})}\BibitemShut {NoStop}%
\bibitem [{\citenamefont {Galiautdinov}(2018)}]{Galiautdinov2018}%
  \BibitemOpen
  \bibfield  {author} {\bibinfo {author} {\bibfnamefont {A.}~\bibnamefont
  {Galiautdinov}},\ }\href@noop {} {\  (\bibinfo {year} {2018})},\ \Eprint
  {http://arxiv.org/abs/arXiv:1805.06877} {arXiv:1805.06877} \BibitemShut
  {NoStop}%
\bibitem [{\citenamefont {Strikis}\ \emph {et~al.}(2019)\citenamefont
  {Strikis}, \citenamefont {Datta},\ and\ \citenamefont {Knee}}]{Strikis2019}%
  \BibitemOpen
  \bibfield  {author} {\bibinfo {author} {\bibfnamefont {A.}~\bibnamefont
  {Strikis}}, \bibinfo {author} {\bibfnamefont {A.}~\bibnamefont {Datta}}, \
  and\ \bibinfo {author} {\bibfnamefont {G.~C.}\ \bibnamefont {Knee}},\ }\href
  {\doibase 10.1103/PhysRevA.99.032328} {\bibfield  {journal} {\bibinfo
  {journal} {Phys. Rev. A}\ }\textbf {\bibinfo {volume} {99}},\ \bibinfo
  {pages} {032328} (\bibinfo {year} {2019})}\BibitemShut {NoStop}%
\bibitem [{\citenamefont {Rol}\ \emph {et~al.}(2019)\citenamefont {Rol},
  \citenamefont {Battistel}, \citenamefont {Malinowski}, \citenamefont
  {Bultink}, \citenamefont {Tarasinski}, \citenamefont {Vollmer}, \citenamefont
  {Haider}, \citenamefont {Muthusubramanian}, \citenamefont {Bruno},
  \citenamefont {Terhal},\ and\ \citenamefont {Dicarlo}}]{Rol2019}%
  \BibitemOpen
  \bibfield  {author} {\bibinfo {author} {\bibfnamefont {M.~A.}\ \bibnamefont
  {Rol}}, \bibinfo {author} {\bibfnamefont {F.}~\bibnamefont {Battistel}},
  \bibinfo {author} {\bibfnamefont {F.~K.}\ \bibnamefont {Malinowski}},
  \bibinfo {author} {\bibfnamefont {C.~C.}\ \bibnamefont {Bultink}}, \bibinfo
  {author} {\bibfnamefont {B.~M.}\ \bibnamefont {Tarasinski}}, \bibinfo
  {author} {\bibfnamefont {R.}~\bibnamefont {Vollmer}}, \bibinfo {author}
  {\bibfnamefont {N.}~\bibnamefont {Haider}}, \bibinfo {author} {\bibfnamefont
  {N.}~\bibnamefont {Muthusubramanian}}, \bibinfo {author} {\bibfnamefont
  {A.}~\bibnamefont {Bruno}}, \bibinfo {author} {\bibfnamefont {B.~M.}\
  \bibnamefont {Terhal}}, \ and\ \bibinfo {author} {\bibfnamefont
  {L.}~\bibnamefont {Dicarlo}},\ }\href@noop {} {\  (\bibinfo {year} {2019})},\
  \Eprint {http://arxiv.org/abs/arXiv:1903.02492v1} {arXiv:1903.02492v1}
  \BibitemShut {NoStop}%
\bibitem [{\citenamefont {Ralph}\ \emph {et~al.}(2005)\citenamefont {Ralph},
  \citenamefont {Hayes},\ and\ \citenamefont {Gilchrist}}]{Ralph2005}%
  \BibitemOpen
  \bibfield  {author} {\bibinfo {author} {\bibfnamefont {T.~C.}\ \bibnamefont
  {Ralph}}, \bibinfo {author} {\bibfnamefont {A.~J.}\ \bibnamefont {Hayes}}, \
  and\ \bibinfo {author} {\bibfnamefont {A.}~\bibnamefont {Gilchrist}},\ }\href
  {\doibase 10.1103/PhysRevLett.95.100501} {\bibfield  {journal} {\bibinfo
  {journal} {Phys. Rev. Lett.}\ }\textbf {\bibinfo {volume} {95}},\ \bibinfo
  {pages} {1} (\bibinfo {year} {2005})}\BibitemShut {NoStop}%
\bibitem [{\citenamefont {Yang}\ \emph {et~al.}(2008)\citenamefont {Yang},
  \citenamefont {Pan}, \citenamefont {Gao}, \citenamefont {Lu}, \citenamefont
  {Zhou},\ and\ \citenamefont {Zhang}}]{Yang2008}%
  \BibitemOpen
  \bibfield  {author} {\bibinfo {author} {\bibfnamefont {T.}~\bibnamefont
  {Yang}}, \bibinfo {author} {\bibfnamefont {J.-W.}\ \bibnamefont {Pan}},
  \bibinfo {author} {\bibfnamefont {W.-B.}\ \bibnamefont {Gao}}, \bibinfo
  {author} {\bibfnamefont {C.-Y.}\ \bibnamefont {Lu}}, \bibinfo {author}
  {\bibfnamefont {X.-Q.}\ \bibnamefont {Zhou}}, \ and\ \bibinfo {author}
  {\bibfnamefont {J.}~\bibnamefont {Zhang}},\ }\href {\doibase
  10.1073/pnas.0800740105} {\bibfield  {journal} {\bibinfo  {journal} {Proc.
  Natl. Acad. Sci.}\ }\textbf {\bibinfo {volume} {105}},\ \bibinfo {pages}
  {11050} (\bibinfo {year} {2008})}\BibitemShut {NoStop}%
\bibitem [{\citenamefont {Mehl}\ \emph {et~al.}(2015)\citenamefont {Mehl},
  \citenamefont {Bluhm},\ and\ \citenamefont {DiVincenzo}}]{Mehl2015}%
  \BibitemOpen
  \bibfield  {author} {\bibinfo {author} {\bibfnamefont {S.}~\bibnamefont
  {Mehl}}, \bibinfo {author} {\bibfnamefont {H.}~\bibnamefont {Bluhm}}, \ and\
  \bibinfo {author} {\bibfnamefont {D.~P.}\ \bibnamefont {DiVincenzo}},\ }\href
  {\doibase 10.1103/PhysRevB.91.085419} {\bibfield  {journal} {\bibinfo
  {journal} {Phys. Rev. B}\ }\textbf {\bibinfo {volume} {91}},\ \bibinfo
  {pages} {085419} (\bibinfo {year} {2015})}\BibitemShut {NoStop}%
\bibitem [{\citenamefont {Andrews}\ \emph {et~al.}(2019)\citenamefont {Andrews}
  \emph {et~al.}}]{Andrews2018}%
  \BibitemOpen
  \bibfield  {author} {\bibinfo {author} {\bibfnamefont {R.~W.}\ \bibnamefont
  {Andrews}} \emph {et~al.},\ }\href {\doibase 10.1038/s41565-019-0500-4}
  {\bibfield  {journal} {\bibinfo  {journal} {Nat. Nanotechnol.}\ } (\bibinfo
  {year} {2019}),\ 10.1038/s41565-019-0500-4}\BibitemShut {NoStop}%
\bibitem [{\citenamefont {Chan}\ and\ \citenamefont {Wang}(2019)}]{Chan2019}%
  \BibitemOpen
  \bibfield  {author} {\bibinfo {author} {\bibfnamefont {G.~X.}\ \bibnamefont
  {Chan}}\ and\ \bibinfo {author} {\bibfnamefont {X.}~\bibnamefont {Wang}},\
  }\href@noop {} {\  (\bibinfo {year} {2019})},\ \Eprint
  {http://arxiv.org/abs/arXiv:1901.04234v1} {arXiv:1901.04234v1} \BibitemShut
  {NoStop}%
\bibitem [{\citenamefont {Grassl}\ \emph {et~al.}(1997)\citenamefont {Grassl},
  \citenamefont {Beth},\ and\ \citenamefont {Pellizzari}}]{Grassl1997}%
  \BibitemOpen
  \bibfield  {author} {\bibinfo {author} {\bibfnamefont {M.}~\bibnamefont
  {Grassl}}, \bibinfo {author} {\bibfnamefont {T.}~\bibnamefont {Beth}}, \ and\
  \bibinfo {author} {\bibfnamefont {T.}~\bibnamefont {Pellizzari}},\ }\href
  {\doibase 10.1103/PhysRevA.56.33} {\bibfield  {journal} {\bibinfo  {journal}
  {Phys. Rev. A}\ }\textbf {\bibinfo {volume} {56}},\ \bibinfo {pages} {33}
  (\bibinfo {year} {1997})}\BibitemShut {NoStop}%
\bibitem [{\citenamefont {{Suchara}}\ \emph {et~al.}(2015)\citenamefont
  {{Suchara}}, \citenamefont {{Cross}},\ and\ \citenamefont
  {{Gambetta}}}]{Suchara2015}%
  \BibitemOpen
  \bibfield  {author} {\bibinfo {author} {\bibfnamefont {M.}~\bibnamefont
  {{Suchara}}}, \bibinfo {author} {\bibfnamefont {A.~W.}\ \bibnamefont
  {{Cross}}}, \ and\ \bibinfo {author} {\bibfnamefont {J.~M.}\ \bibnamefont
  {{Gambetta}}},\ }in\ \href {\doibase 10.1109/ISIT.2015.7282629} {\emph
  {\bibinfo {booktitle} {2015 IEEE International Symposium on Information
  Theory (ISIT)}}}\ (\bibinfo {year} {2015})\ pp.\ \bibinfo {pages}
  {1119--1123}\BibitemShut {NoStop}%
\bibitem [{\citenamefont {Delfosse}\ and\ \citenamefont
  {Z{\'{e}}mor}(2017)}]{Delfosse2017a}%
  \BibitemOpen
  \bibfield  {author} {\bibinfo {author} {\bibfnamefont {N.}~\bibnamefont
  {Delfosse}}\ and\ \bibinfo {author} {\bibfnamefont {G.}~\bibnamefont
  {Z{\'{e}}mor}},\ }\href@noop {} {\  (\bibinfo {year} {2017})},\ \Eprint
  {http://arxiv.org/abs/arXiv:1703.01517} {arXiv:1703.01517} \BibitemShut
  {NoStop}%
\bibitem [{\citenamefont {Delfosse}\ and\ \citenamefont
  {Nickerson}(2017)}]{Delfosse2017b}%
  \BibitemOpen
  \bibfield  {author} {\bibinfo {author} {\bibfnamefont {N.}~\bibnamefont
  {Delfosse}}\ and\ \bibinfo {author} {\bibfnamefont {N.~H.}\ \bibnamefont
  {Nickerson}},\ }\href@noop {} {\  (\bibinfo {year} {2017})},\ \Eprint
  {http://arxiv.org/abs/arXiv:1709.06218v1} {arXiv:1709.06218v1} \BibitemShut
  {NoStop}%
\bibitem [{\citenamefont {Stace}\ \emph {et~al.}(2009)\citenamefont {Stace},
  \citenamefont {Barrett},\ and\ \citenamefont {Doherty}}]{Stace2009}%
  \BibitemOpen
  \bibfield  {author} {\bibinfo {author} {\bibfnamefont {T.~M.}\ \bibnamefont
  {Stace}}, \bibinfo {author} {\bibfnamefont {S.~D.}\ \bibnamefont {Barrett}},
  \ and\ \bibinfo {author} {\bibfnamefont {A.~C.}\ \bibnamefont {Doherty}},\
  }\href {\doibase 10.1103/PhysRevLett.102.200501} {\bibfield  {journal}
  {\bibinfo  {journal} {Phys. Rev. Lett.}\ }\textbf {\bibinfo {volume} {102}},\
  \bibinfo {pages} {200501} (\bibinfo {year} {2009})}\BibitemShut {NoStop}%
\bibitem [{\citenamefont {Stace}\ and\ \citenamefont
  {Barrett}(2010)}]{Stace2010}%
  \BibitemOpen
  \bibfield  {author} {\bibinfo {author} {\bibfnamefont {T.~M.}\ \bibnamefont
  {Stace}}\ and\ \bibinfo {author} {\bibfnamefont {S.~D.}\ \bibnamefont
  {Barrett}},\ }\href {\doibase 10.1103/PhysRevA.81.022317} {\bibfield
  {journal} {\bibinfo  {journal} {Phys. Rev. A}\ }\textbf {\bibinfo {volume}
  {81}},\ \bibinfo {pages} {022317} (\bibinfo {year} {2010})}\BibitemShut
  {NoStop}%
\bibitem [{\citenamefont {Vodola}\ \emph {et~al.}(2018)\citenamefont {Vodola},
  \citenamefont {Amaro}, \citenamefont {Martin-Delgado},\ and\ \citenamefont
  {M\"uller}}]{Vodola2018}%
  \BibitemOpen
  \bibfield  {author} {\bibinfo {author} {\bibfnamefont {D.}~\bibnamefont
  {Vodola}}, \bibinfo {author} {\bibfnamefont {D.}~\bibnamefont {Amaro}},
  \bibinfo {author} {\bibfnamefont {M.~A.}\ \bibnamefont {Martin-Delgado}}, \
  and\ \bibinfo {author} {\bibfnamefont {M.}~\bibnamefont {M\"uller}},\ }\href
  {\doibase 10.1103/PhysRevLett.121.060501} {\bibfield  {journal} {\bibinfo
  {journal} {Phys. Rev. Lett.}\ }\textbf {\bibinfo {volume} {121}},\ \bibinfo
  {pages} {060501} (\bibinfo {year} {2018})}\BibitemShut {NoStop}%
\bibitem [{\citenamefont {Aloshious}\ \emph {et~al.}(2018)\citenamefont
  {Aloshious}, \citenamefont {Bhagoji},\ and\ \citenamefont
  {Sarvepalli}}]{Aloshious2018}%
  \BibitemOpen
  \bibfield  {author} {\bibinfo {author} {\bibfnamefont {A.~B.}\ \bibnamefont
  {Aloshious}}, \bibinfo {author} {\bibfnamefont {A.~N.}\ \bibnamefont
  {Bhagoji}}, \ and\ \bibinfo {author} {\bibfnamefont {P.~K.}\ \bibnamefont
  {Sarvepalli}},\ }\href@noop {} {\  (\bibinfo {year} {2018})},\ \Eprint
  {http://arxiv.org/abs/arXiv:1804.00866} {arXiv:1804.00866} \BibitemShut
  {NoStop}%
\bibitem [{\citenamefont {{Gottesman}}(1997)}]{Gottesman1997}%
  \BibitemOpen
  \bibfield  {author} {\bibinfo {author} {\bibfnamefont {D.}~\bibnamefont
  {{Gottesman}}},\ }\emph {\bibinfo {title} {{Stabilizer codes and quantum
  error correction}}},\ \href {https://arxiv.org/abs/quant-ph/9705052} {Ph.D.
  thesis},\ \bibinfo  {school} {California Institute of Technology} (\bibinfo
  {year} {1997})\BibitemShut {NoStop}%
\bibitem [{\citenamefont {Stauffer}(1985)}]{Stauffer1985}%
  \BibitemOpen
  \bibfield  {author} {\bibinfo {author} {\bibfnamefont {D.}~\bibnamefont
  {Stauffer}},\ }\href@noop {} {\emph {\bibinfo {title} {{Introduction to
  percolation theory}}}}\ (\bibinfo  {publisher} {Taylor {\&} Francis},\
  \bibinfo {address} {Lodon},\ \bibinfo {year} {1985})\BibitemShut {NoStop}%
\bibitem [{\citenamefont {Feng}\ \emph {et~al.}(2008)\citenamefont {Feng},
  \citenamefont {Deng},\ and\ \citenamefont {Bl\"ote}}]{Feng2008}%
  \BibitemOpen
  \bibfield  {author} {\bibinfo {author} {\bibfnamefont {X.}~\bibnamefont
  {Feng}}, \bibinfo {author} {\bibfnamefont {Y.}~\bibnamefont {Deng}}, \ and\
  \bibinfo {author} {\bibfnamefont {H.~W.~J.}\ \bibnamefont {Bl\"ote}},\ }\href
  {\doibase 10.1103/PhysRevE.78.031136} {\bibfield  {journal} {\bibinfo
  {journal} {Phys. Rev. E}\ }\textbf {\bibinfo {volume} {78}},\ \bibinfo
  {pages} {031136} (\bibinfo {year} {2008})}\BibitemShut {NoStop}%
\end{thebibliography}%
\bibliographystyle{apsrev4-1}

\end{document}